\documentclass[12pt, twoside, here]{article}
\usepackage{epsf}
\usepackage{times,colordvi,amsmath,epsfig,float,color,multicol}
\usepackage{graphics}
\usepackage{hhline}
\usepackage[large]{subfigure}
\usepackage[latin1]{inputenc}
\usepackage{rotating}

\newtheorem{assumption}{\sc Assumption}

\title{A Model of Teneral Dehydration in {\em Glossina}}

\author{S. J. Childs \\ \\ {\small\em Department of Mathematics and Applied Mathematics, University of the Free State,} \\ {\small\em P.O. Box 339, Bloemfontein, 9300, South Africa.} \\ {\small\em Tel: +27 51 4013386 \ \ Email: simonjohnchilds@gmail.com}}

\renewcommand{\thefootnote}{\fnsymbol{footnote}}
\date{Acta Tropica, 131: 79--91, 2014}

\begin{document}

\maketitle
\renewcommand{\thefootnote}{\arabic{footnote}}

\begin{abstract}
\noindent {\em The results of a long-established investigation into teneral
transpiration are used as a rudimentary data set. These data are not complete in
that all are at 25~$^\circ\mathrm{C}$ and the temperature-dependence cannot,
therefore be resolved. An allowance is, nonetheless, made for the outstanding
temperature-dependent data. The data are generalised to all humidities, levels
of activity and, in theory, temperatures, by invoking the property of
multiplicative separability. In this way a formulation, which is a very simple,
first order, ordinary differential equation, is devised. The model is extended
to include a variety of Glossina species by resorting to their relative, resting
water loss rates in dry air. The calculated, total water loss is converted to
the relevant humidity, at 24~$^\circ\mathrm{C}$, that which produced an
equivalent water loss in the pupa, in order to exploit an adaption of an
established survival relationship. The resulting computational model calculates
total, teneral water loss, consequent mortality and adult recruitment.
Surprisingly, the postulated race against time, to feed, applies more to the
mesophilic and xerophilic species, in that increasing order. So much so that it
is reasonable to conclude that, should Glossina brevipalpis survive the pupal
phase, it will almost certainly survive to locate a host, without there being
any significant prospect of death from dehydration. With the conclusion of this
work comes the revelation that the classification of species as hygrophilic,
mesophilic and xerophilic is largely true only in so much as their third and
fourth instars are and, possibly, the hours shortly before eclosion.} 
\end{abstract}

Keywords: teneral water loss; transpiration; dehydration; mortality; tsetse; {\em glossina}. 

\section{Introduction}

The teneral stage of the tsetse fly is defined to commence immediately after
eclosion and it terminates with the taking of the first blood-meal. Water loss
continues after eclosion and dehydration becomes progressively, more critical,
up until the moment the teneral fly has its first meal. Once fed, it is
essentially no longer in jeopardy. The severely depleted, pupal reserves are
replenished and, from then on, dehydration assumes a far lesser importance. The
adult fly is far better equipped to fend for itself and, to a certain extent, is
a master of its own destiny. It can avoid dehydration through behavioural
strategies, for example, by modifying its level of activity, by temporarily
retreating to deep shade, or by locating an host on which to feed; all
activities that the severely depleted reserves of the teneral fly may not allow
enough time for. Work on adult flies, by Hargrove\nocite{Hargrove3} (2004), is
strongly supportive of such reasoning in that humidity was found to be
insignificant with regard to adult mortality. Dehydration is therefore a
phenomenon usually only associated with the pupal and teneral stages in tsetse.
The ultimate toll on a given cohort is cumulative and, likely, best assessed in
terms of the proportion of original larvae which still have sufficient reserves
to achieve their first feed, as tenerals. 

Since combined dehydration and fat loss are thought to culminate in massive
teneral mortality (Hargrove\nocite{Hargrove1}, 1990), pupal and teneral
mortality rates are crucial in deciding the viability of any tsetse population.
The vastly different dynamics of water loss during the pupal and teneral phases,
however, afford both the status of topics in their own right. It seems likely
that the high mortality which usually characterises the teneral phase is
essentially determined during the pupal phase. While teneral water loss rates
are generally several times higher than pupal rates, it can be argued that pupal
rates prevail many times longer  (comparing the
Bursell\nocite{Bursell1}\nocite{Bursell2}, 1958 and 1959, data). To give some
idea of the relative importance, while teneral water loss rates are probably
around 40 times sensu strictu pupal-stage rates and around seven times the
unprotected rates which prevail prior to and immediately following the sensu
strictu pupal-stage, the unprotected rates generally prevail six times longer
than teneral rates and sensu strictu pupal-stage rates 24 times longer. Thus,
any teneral that dies of dehydration could be said, very likely, to have died as
a result of pupal water loss. Water loss during the pupal phase can decide the
fate of the teneral and one possible criticism of the Bursell\nocite{Bursell2}
(1959) work is that it doesn't take the state of the inherited, pupal reserves
into account enough. 

A general model of teneral water loss is developed in the same vein as the pupal
water loss model of Childs\nocite{Childs1}\nocite{Childs2} (2013 and 2009). It
is largely based on the investigations of one experimentalist
(Bursell\nocite{Bursell1}\nocite{Bursell2}, 1959 and 1958) and it otherwise
relies heavily on the pupal dehydration model for its initial values. The main
challenge to exploiting Bursell's results for the purposes of a computational
model, could be said to be in three, very specific respects: A function for
transpiration, extending the {\em Glossina morsitans}-based, teneral model to
the rest of the {\em Glossina} genus, then formulating a satisfactory criterion
for survival that is dependent on total water loss. In the latter instance, a
surprisingly simple solution is found to lie in the form of the pupal emergence
data and the challenge then becomes one of utilizing the dependence of pupal
survival on humidity, at 24~$^\circ\mathrm{C}$, when only the cumulative water
loss for the teneral has been calculated. Transpiration is activity-dependent in
the case of the teneral and an allowance for the future acquisition of
temperature-dependent data must be made in generalising transpiration to all
humidities, activity levels and temperatures. The Bursell\nocite{Bursell2}
(1959) teneral work was all carried out at 25 $\pm$ 0.8~$^\circ\mathrm{C}$. In
the pupal case, the {\em G. morsitans} model was extended to other species on
the basis of the puparium's surface area and its transpiration rate per unit of
surface area, a strategy which predicted the critical water losses of all
species with remarkable success in \mbox{Childs\nocite{Childs1}\nocite{Childs2}
(2013 and 2009)}. Bursell\nocite{Bursell2} (1959) presents a convincing argument
that, within a single species, recording transpiration in units of residual dry
mass is a superior means to that utilizing units of total mass per surface area.
It is claimed that such a measurement is more resilient to phenotypic
plasticity. It is based on such data that the {\em G. morsitans}-based model
must necessarily be extended to other species. Extending a {\em G.
morsitans}-based model to other species is work that can best be described as
exploratory, however, the success of the same approach in the pupal model is
cause for optimism. The results at the end make it an interesting and
justifiable exercise, nonetheless. This model takes no account of the vagaries
of phenotypic plasticity although there is no reason why relevant data would not
facilitate the incorporation of such detail.

The final formulation, hence solution to the problem, is predicated on five
major assumptions which are explicitly stated and explored. Another is taken for
granted. It is assumed that the Bursell\nocite{Bursell2} (1958 and 1959)
investigations are comprehensive, to the extent that they encapsulate all
salient aspects of pupal and teneral water loss. The problem of teneral
transpiration is then reduced to a first order, ordinary differential equation
for water loss. Although this equation is in itself extremely simple, the other
equations, which constitute the combined pupal and teneral scheme, are both
numerous and voluminous and there are issues pertaining to differentiability and
continuity. This fact and the anticipated accuracy of such models render
preferred integration schemes, such as the fourth-order-accurate
Runge-Kutta-Fehlberg (RKF45) method, slightly impractical. Since the problem
is not intractably large from a computational point of view, expedience takes
precedence over taste and the more pedestrian midpoint rule is the preferred
integration technique, in keeping with the pupal-stage model. It is in this way
that the resulting problem is transformed from a mathematical one, to a
computational one. 

The goals and broader applications of this work are threefold. In order of
priority, the first is the completion of the most challenging compartment of an
early mortality model, the second is habitat assessment, while the third is a
better comprehension of tsetse biology, particularly the
Bursell\nocite{Bursell1}\nocite{Bursell2} (1958 and 1959) endeavours. Most of
the experimental work needed for a model of early stage mortality has long been
complete. The main causes of mortality could be summed up as dehydration, fat
loss, predation and parasitism. The relationship between pupal fat loss and
temperature has been extensively studied by Bursell\nocite{Bursell3} (1960) and
Phelps\nocite{Phelps1} (1973). A cursory inspection of that work suggests that a
few data points pertaining to teneral fat consumption's dependence on activity
(at either fixed or variable temperature) are required. Quantitative work linking predation and parasitism to the density at
pupal sites has been carried out by Rogers and
Randolph\nocite{RogersAndRandolph1} (1990), although the topic can almost
certainly be predicted to require some stochastic treatment. The completion of
this work could be supposed to leave the way open to a comprehensive model of
early mortality, based either on a joint probability density function, or, more
likely, a Markov chain. Once a model of early stage mortality is completed,
matters should become a lot simpler. Adult mortality is lower
(Hargrove\nocite{Hargrove1}, 1990) and thought to be trivial, likely a simple,
linear dependence on temperature with population density becoming relevant at
its higher levels. 

The question of habitat assessment is a topic of intense interest to
entomologists and parasitologists. Although environmental degradation militates
against the tsetse fly ever making any significant return to its former status,
health officials might have become over-reliant on work such as \mbox{Ford and
Katondo\nocite{FordAndKatondo} (1977)} and the implications of climate change
for traditional habitat (as well as other, geopolitical factors) could find them
complacent. The presently perceived incursion of {\em G. austeni} into
previously unrecognised, South African habitat (Hendrickx\nocite{Hendrickx},
2007), for example, requires an explanation. Accurate determination of a greatly
more confined pupal and teneral habitat, as well as early mortality, will
facilitate a greatly more effective application of the various kinds of control
measures contemplated by an integrated approach to pest management (Barclay and
Vreysen\nocite{BarclayAndVreysen}, 2010), aerial spraying being the possible
exception (Childs\nocite{Childs5}\nocite{Childs4}, 2013 and 2011). Habitat
assessment and tsetse biology are, of course, intricately entwined. The
implications of tsetse biology for the habitat assessment of hygrophilic species
in general, as well as South Africa's two, extant tsetse species, {\em G.
brevipalpis} and {\em G. austeni}, are elucidated in this work. Certainly there
are surprises in store so far as to what truly sets hygrophilic species apart
from their mesophilic and xerophilic counterparts.

\section{Generalising the Transpiration Data to a Function of Humidity, Activity and Temperature}

The transpiration rate resorted to in this work is measured in residual dry
masses per hour, rather than as a pecentage of the same which was the preferred
choice of Bursell\nocite{Bursell2} (1959). All other units conform to those of
Bursell\nocite{Bursell2} (1959). Both the level of activity and the relative
humidity used are percentages and the temperature is in degrees centigrade.
Bursell \nocite{Bursell2} (1959) obtained one set of transpiration data for
variable activity (at 0\% $\mathrm{r.h.}$ and 80\% $\mathrm{r.h.}$) and another
for variable humidity (at 0\% acitivity and 30\% activity) during his
investigations into teneral water loss (Fig. 1 of Bursell\nocite{Bursell2},
1959). Both data sets were measured at a temperature of 25 $\pm$ 0.8
$^\circ\mathrm{C}$. No temperature-dependent data are presently known to exist,
however, this need not preclude one from making provision for such data coming
into existence at some stage in the future. The question as to how one
generalises these and as yet unknown data to all temperatures, humidities and
levels of activity therefore arises. The prospect of some, temperature-dependent
data set coming into existence in the future lends itself favourably to an
assumption of multiplicative seperability. 
\begin{assumption} \label{assumption1}
{\bf \em Transpiration rate is a multiplicatively, separable function of
activity and temperature. That is, if $dk/dt$ is the
transpiration rate, then there exist two functions $\phi$ and $\theta$,
dependent exclusively on activity and temperature respectively, so that} 
\begin{eqnarray} \label{1}
\frac{dk}{dt}(h,a,T) &=& \phi(h,a) \ \theta(h,T),
\end{eqnarray} 
{\bf \em in which $a$ denotes activity, $T$ the temperature and $h$ is the humidity.} 
\end{assumption} 
Such an assumption supposes that transpiration's dependence on humidity at one
temperature, is simply a temperature-dependent multiple of that same dependence
at another temperature i.e. that there is no coupling of the independent
variables. Simple relationships, such as the transpiration rate depicted in
Bursell\nocite{Bursell2} (1959), often lend themselves favourably to such an
assumption. This is especially given that one would anticipate any,
temperature-dependent transpiration function to be simple, smooth, continuous
and monotonic; very likely pure, exponential in dry air, as is the case during
the pupal stage (Bursell\nocite{Bursell2}, 1958). The asymptotic errors in
Table \ref{coefficients}, for example, can easily be interpreted to justify the
omission of both quadratic terms in \mbox{Eq. \ref{fitToData}} and, therefore,
to replace it with a separable function, instead. Thus, in the very
likely event that water loss rates are not multiplicatively separable,
multiplicative separability should not be a bad substitute. 

\subsection{The Dependence of Transpiration on Activity and Humidity}

The transpiration rate during the teneral phase is approximately steady. It
differs little from the adult rate, which is marginally lower (comparing Figures
1A and 1B of Bursell\nocite{Bursell2}, 1959). An humidity and activity
dependence in keeping with the multiplicative separability assumption is
recognized and the relevant data is that published in Figs. 1A, 1C and 1D of
Bursell\nocite{Bursell2} (1959). The figures can be interpreted as transections
through a surface which intersect. They suggest a very simple surface, one which
appears to be of no higher order than bi-quadratic, by inspection. It was
therefore decided to fit a bi-quadratic surface,
\begin{eqnarray} \label{fitToData}
{\frac{dk}{dt}}(h,a,25) &=& c_1 + c_2h + c_3a + c_4ha + c_5h^2 + c_6a^2, 
\end{eqnarray} 
to the transpiration data using the method of least squares and the values of
the coefficients, thus obtained, are tabulated in Table \ref{coefficients}. 
\begin{table}[H]
    \begin{center}
\begin{tabular}{c|l l}  
&  &  \\
 coefficient \ & ~ ~ ~ ~ ~ value \ & \ ~ asymptotic standard error \\
&  & \\ \hline 
&  & \\
$c_1$ \ & ~ \ $2.49378 \times 10^{-2}$ & ~ ~ ~ ~ ~ ~ $\pm 0.001011$ \\
$c_2$ \ & $- 1.18516 \times 10^{-4}$ & ~ ~ ~ ~ ~ ~ $\pm 5.351 \times 10^{-5}$ \\
$c_3$ \ & ~ \ $5.5411 \times 10^{-4}$ & ~ ~ ~ ~ ~ ~ $\pm 0.0001082$ \\
$c_4$ \ & $- 3.86355 \times 10^{-6}$ & ~ ~ ~ ~ ~ ~ $\pm 7.921 \times 10^{-7}$ \\
$c_5$ \ & $- 9.44705 \times 10^{-7}$ & ~ ~ ~ ~ ~ ~ $\pm 5.874 \times 10^{-7}$ \\
$c_6$ \ & $- 2.07566 \times 10^{-6}$ & ~ ~ ~ ~ ~ ~ $\pm 2.741 \times 10^{-6}$ \\
&  & \\
\end{tabular}
\caption{Coefficients for a bi-quadratic fit to the Bursell\nocite{Bursell2} (1959) transpiration data. The sum of squares of residuals was $1.03066 \times 10^{-4}$.}
\label{coefficients}
    \end{center}
\end{table}
Note that the asymptotic standard errors in Table \ref{coefficients} appear to
deny any justification for a term which is purely quadratic in activity,
although including such a term does result in a marginally improved sum of the
squares of the residuals. That selfsame observation might also be said to
pertain to the term which is purely quadratic in humidity. Yet a single glance
at Fig. 1C of Bursell\protect\nocite{Bursell2} (1959) will leave little doubt
that that nonlinear term indeed exists. A subtle distinction between a
term being inappropriate and the quality and quantity of the data was therefore
invoked to retain all the nonlinear terms. Fortunately, the contentious
decision, supported by somewhat heuristic arguments, is of no consequence for
all realistic levels of activity. Only for levels of activity above 50\% is
there a slight difference in the transpiration rates predicted when omitting the
somewhat controversial quadratic activity term. 

\begin{figure}[H]
    \begin{center}
\includegraphics[height=11cm, clip = true]{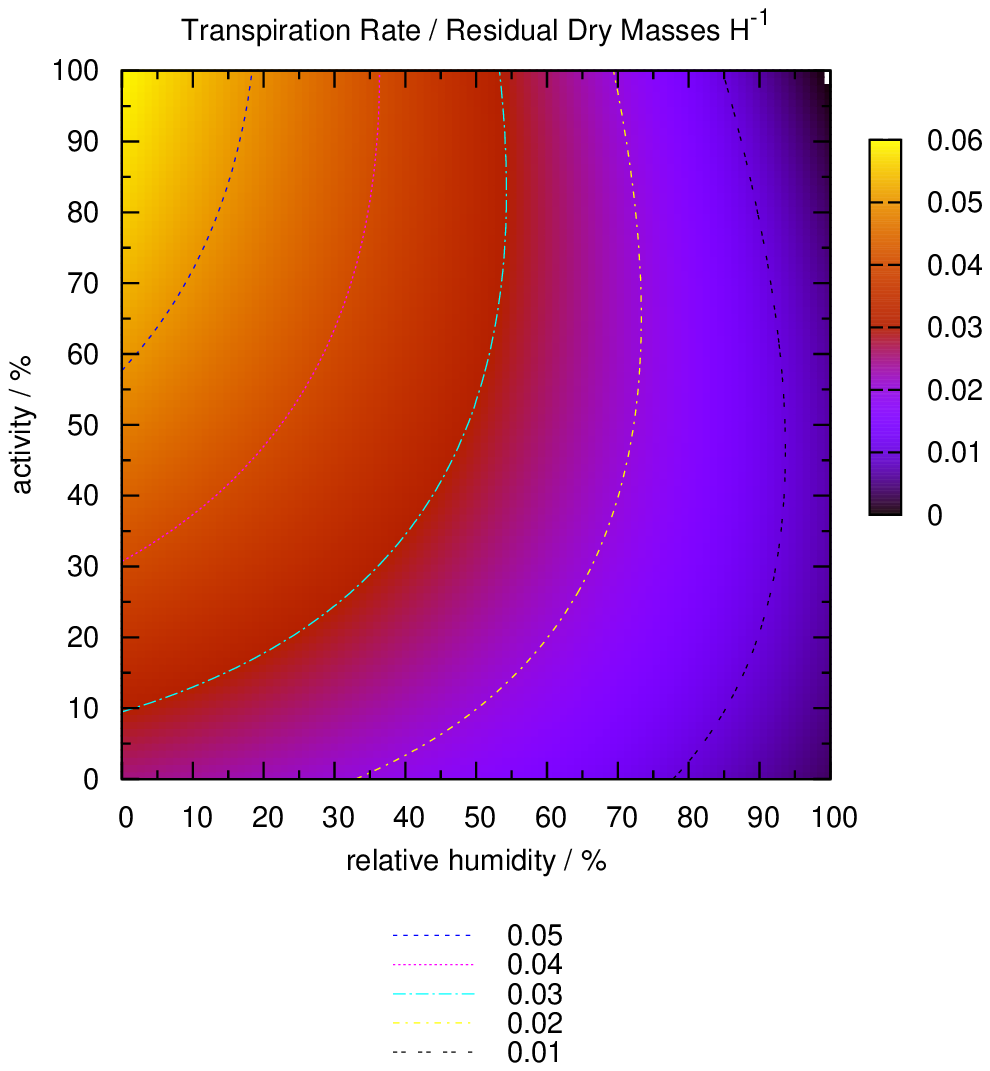}
\caption{The fit to the Bursell\protect\nocite{Bursell2} (1959) Figs. 1A, 1C and 1D data.} \label{bursellTeneral}
   \end{center}
\end{figure}

The resulting bi-quadratic fit is in {\em G. morsitans}-teneral, residual dry masses per hour and is depicted in \mbox{Fig. \ref{bursellTeneral}}. That surface is nonetheless a surface of transpiration when, instead, only the humidity and activity dependence are sought. The multiplicative separability assumption facilitates the following manipulation,
\begin{eqnarray} \label{phi}
\phi(h,a) &=& \frac{c_1 + c_2h + c_3a + c_4ha + c_5h^2 + c_6a^2}{\theta(h,25)},
\end{eqnarray} 
and thus, an expression for the combined humidity--activity dependence is obtained. It is in this way that the model facilitates the future acquisition of temperature-dependent data.

\subsection{The Dependence of Transpiration on Temperature}

All Bursell \nocite{Bursell2} (1959) data sets were recorded at a constant
temperature of 25 $\pm$ 0.8~$^\circ\mathrm{C}$. They provide no clue as to the
dependence of transpiration on temperature. For this reason further laboratory
measurements are needed. Measuring activity levels is a tedious task which can
probably be avoided if cognizance is taken of the work of Brady\nocite{Brady1}
(1972). There is no activity at night for flies kept continuously in the dark
(Brady\nocite{Brady1}, 1972). 

\section{Extending the Model to Other Species}

It is generally suspected that the {\em Glossina} genus derives from a common,
tropical, rain-forest dwelling ancestor, adjusted to moist, warm climates
(Glasgow\nocite{Glasgow1}, 1963). There can be no doubt that the water reserve
is a limiting factor to the pupa (Bursell\nocite{Bursell1}, 1958) and the young,
teneral fly has not yet had an opportunity to replenish that reserve.
Dehydration is therefore a challenge, if not the major threat to the teneral
fly, when one considers the significance Rogers and
Robinson\nocite{RogersAndRobinson} (2004) attach to cold cloud duration
(rainfall). One might therefore presume that the tenerals, in all tsetse
species, actively pursue the same strategy to minimise water loss for the
majority of modern habitats and have mechanisms preventative of dehydration. 

Adult transpiration rates in dry air and at room temperature have been measured
as a percentage of the residual dry mass for a number of species and are quoted
in Table II of Bursell\nocite{Bursell2} (1959). The residual dry masses,
themselves, can be recovered from Fig. 2 of Bursell\nocite{Bursell2} (1959).
That same Bursell\nocite{Bursell2} (1959) endeavour suggests adult transpiration
rates are not significantly different from those of the teneral and should
suffice for the teneral model. Transpiration during the teneral phase can
therefore also be assumed to be steady and the observation that transpiration in
adults is only minutely lower than in tenerals (Bursell\nocite{Bursell2}, 1959)
would seem to vindicate an assumption that the decrease takes place over a
relatively long period. The resting transpiration rate per fly, in dry air at
room temperature, can therefore be calculated from this data. It is the relevant
datum multiplied by the residual dry mass, divided by 100. Under such conditions
a dimensionless, species conversion factor for teneral transpiration rates can
be defined as the ratio of the water loss rates of the different species. That
is,
\begin{eqnarray} \label{11}
\delta_{\mbox{\scriptsize species}} &=& \frac{ p_{\mbox{\scriptsize species}} }{ p_{\mbox{\scriptsize morsitans}} } \times \frac{ m_{\mbox{\scriptsize species}} }{ m_{\mbox{\scriptsize morsitans}} }, \nonumber 
\end{eqnarray}
in which $p_{\mbox{\scriptsize morsitans}}$ and $p_{\mbox{\scriptsize species}}$
are the resting water loss rates, in dry air, for {\em G. morsitans} and the
species in question, respectively; $m_{\mbox{\scriptsize morsitans}}$ and
$m_{\mbox{\scriptsize species}}$ are the residual dry masses of \mbox{{\em G.
morsitans}} and the species in question, respectively. Actual values of
$\delta_{\mbox{\scriptsize species}}$ for eight different species are tabulated
in Table \ref{modelConversionFactors}.
\begin{table}[H]
    \begin{center}
\begin{tabular}{l l l}  
&  &  \\
~ Group \ & \ ~ Species \ & \ ~ ~ $\delta$ \\
&  & \\ \hline 
&  & \\
{\em morsitans} \ & \ {\em austeni} \ & \ 0.952 \\ 
 \ & \ {\em morsitans} & \ 1.00 \\ 
 \ & \ {\em pallidipes} & \ 1.43 \\ 
 \ & \ {\em swynnertoni} \ & \ 1.10 \\ 
& & \\
{\em palpalis} \ & \ {\em palpalis} \ & \ 1.19 \\ 
&  & \\ 
{\em fusca} \ & \ {\em brevipalpis} \ & \ 2.10 \\ 
 \ & \ {\em fuscipleuris} \ & \ 2.26 \\ 
 \ & \ {\em longipennis} \ & \ 3.76 \\ 
&  & \\
\end{tabular}
\caption{Species conversion factors based on resting, adult tranpsiration rates and the residual, dry masses of newly-eclosed tenerals (Table II and
Fig. 2 respectively, of Bursell\nocite{Bursell2}, 1959).}
\label{modelConversionFactors}
    \end{center}
\end{table}
There is good reason to believe that these data have the potential for the
extrapolation of the {\em G. morsitans} model, to other species, in a more
general context. These factors could enable the teneral transpiration rate for
another species to be extrapolated from {\em G. morsitans} values, based on the following assumption. 
\begin{assumption} \label{assumption3} 
{\bf \em The relative, resting transpiration rates, in dry air at 25~$^\circ\mathrm{C}$, can be used to extrapolate the hypothetical water loss, in one species, into the water loss of another species, for any given activity level and any set of environmental conditions.}
\end{assumption} 
Assumption \ref{assumption3} certainly appears to be the most tenuous and one
cannot help questioning its validity, wondering whether such a simplistic
approach will work. Although no variation in rates with pronounced changes in
temperature and humidity is indicated, it is of some comfort that the conversion
of the {\em G. morsitans} model to other species involves relative rates. The
factors are a comparative ratio of the species-specific,
transpiration-rate-per-fly data. Certainly there was surprisingly strong
evidence to suggest that the analogous assumption worked from the behavioural
point of view, in the pupal case. In the pupal scenario, the relative surface
areas and the relative permeability of the membranes, per unit area of membrane,
were used to convert the {\em G. morsitans} water loss to those of another
species (Childs\nocite{Childs1}\nocite{Childs2}, 2013 and 2009). The strategy
for the teneral is the same in that transpiration data are converted to a
per-fly value for each species, in order for a relative value to be obtained. In
the case of the teneral fly, it is an inescapable fact that there is less scope
for biologically different strategies among the tenerals of the different
species, than there is for the pupae from which they derive. Transpiration is
steady. There are no unprotected third and fourth instars, followed by a
protracted transition to a protected, sensu strictu pupal stage etc. Allowance
has been made to accommodate the most basic behavioural change, namely the level
of physical activity. Of course, although quantifiable, predicting that activity
level might prove to be as difficult as predicting the temperature and humidity
of ``deep shade'', although activity does appear to be driven by dehydration
(Bursell\nocite{Bursell2}, 1959). Both these problems are, nonetheless,
irrelevant to Assumption \ref{assumption3}.

It should be noted that the {\em G brevipalpis} data in Fig. 2 of
Bursell\nocite{Bursell2} (1959) appear to be at odds with Table III of that same
publication. Although the {\em G. brevipalpis} residual dry weight is expected
to be approximately in keeping with that of {\em G. fuscipleuris}, it was
reasoned, firstly, that {\em G. brevipalpis} has a higher water loss rate prior
to eclosion and, secondly, that a single typographical error is more likely than
four, incorrectly plotted coordinates. The relevant, Bursell\nocite{Bursell2}
(1959) Fig. 2 value for {\em G brevipalpis} was accordingly selected over that
in Table III of the same reference.

\section{The Resulting Model for Teneral Water Loss}

The teneral transpiration rate can be formulated by collecting together all prior derivation and the assumptions to give rise to a single governing equation. The rate formula which follows constitutes a first order, ordinary differential equation.  

\subsection{The Governing Equation}

Generalising Eqs. \ref{1} and \ref{phi} to all species leads to
\begin{eqnarray} \label{17}
\frac{dk}{dt} &=& \left[ c_1 + c_2h + c_3a + c_4ha + c_5h^2 + c_6a^2 \right] \ \frac{ \theta(h, T) }{ \theta(h, 25) } \ \frac{ p_{\mbox{\scriptsize species}} }{ p_{\mbox{\scriptsize morsitans}} } \frac{ m_{\mbox{\scriptsize species}} }{ m_{\mbox{\scriptsize morsitans}} }
\end{eqnarray}
in units of {\em G. morsitans}-teneral, residual dry masses per hour
(\mbox{5.43 $\mathrm{mg}$ $\mathrm{h}^{-1}$}) and in which the coefficients are those from Table \ref{coefficients}. At 25~$^\circ\mathrm{C}$ the
fraction involving the unknown function, $\frac{ \theta(h, T) }{ \theta(h, 25)
}$, is unity.

\subsection{Numerical Integration}

The above rate formula constitutes a first order, ordinary differential
equation. A fourth order accurate Runge-Kutta-Fehlberg method (RKF45) would
normally be the preferred method of integration as it combines high accuracy
with error control. A scheme in keeping with that of the pupal model it is used
in conjunction with is, however, more convenient. 

The midpoint rule is usually considered distasteful from the point of view of
its error. The local error per step, of length ${\Delta t}$, is $\mathop{\rm
O}({\Delta t}^3)$. Since the required number of steps is proportional to
$\frac{1}{{\Delta t}}$, the global error is $\mathop{\rm O}({\Delta t}^2)$. This
is indeed primitive. The real strength of the midpoint rule and other low order
methods lies, however, in their robustness at discontinuities and points of
non-differentiability; something which is indeed relevant to the pupal models of
Childs\nocite{Childs1}\nocite{Childs2} (2013 and 2009). The maximum, additional
error introduced at such points is of nearly the same order as the method's
global error. The same cannot be said for the higher order methods. The handicap
of a poor error is easily overcome computationally by using a small step length,
although the midpoint rule can never be considered adaptive in the same way as
the Runge-Kutta-Fehlberg method.

When considering the original pupal material used, that an engineering-type
accuracy is anticipated from the model and a host of other factors, two
significant figures are more than what are sought. Since the problem is not
intractably large, expedience takes precedence over taste and the more
pedestrian midpoint rule is considered the appropriate choice. 

\section{The Effect of Dehydration on Teneral Survival and Adult Recruitment} \label{adulthood}

\begin{figure}[H]
    \begin{center}
\includegraphics[height=14cm, angle=-90, clip = true]{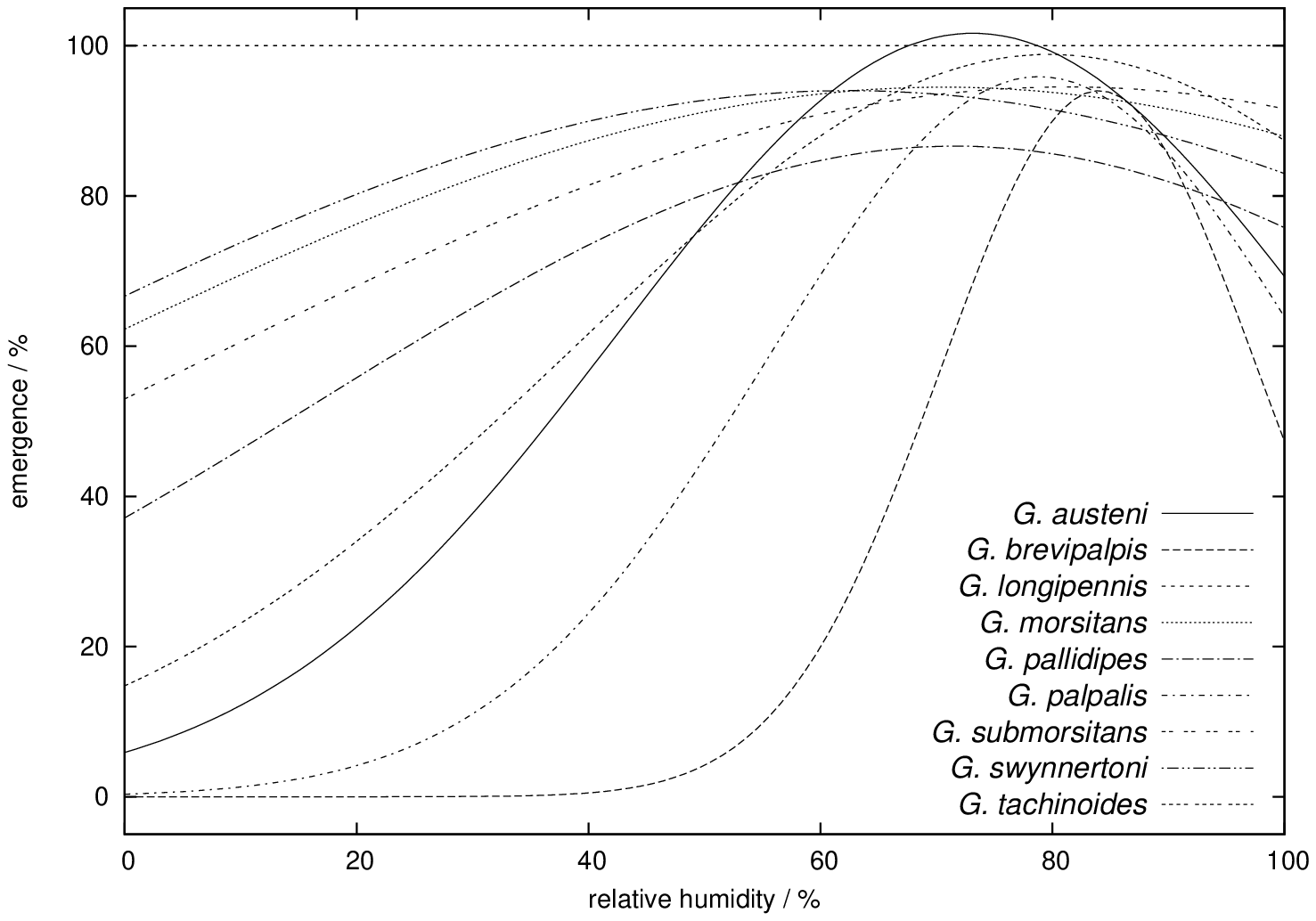}
\caption{Percentage emergence data (Bursell, 1958, and Buxton and Lewis, 1934)
modelled as a Gaussian curve (Childs, 20013 and 2009) for a variety of species.
All are at 24~$^\circ\mathrm{C}$, except {\em G. tachinoides}
(30~$^\circ\mathrm{C}$). {\em G. longipennis} is the single exception (a
straight line had to be fitted to the only two data points).}
\label{allSpeciesTogether}
   \end{center}
\end{figure}

It is logical to assume that the same level of depletion that was fatal to the
pupa before eclosion, only a few hours before, will likewise be fatal to the
newly-eclosed, teneral fly. Since the newly-eclosed, teneral has not yet fed,
one can reasonably argue that it inherits the pupa's remaining reserves and will
die off in the same cumulative proportions, as dehydration becomes progressively
more critical. The failure of the pupa to eclode is simply a measure of
mortality. It is a symptom of the same water reserve which the teneral inherits,
becoming depleted to critical levels. 
\begin{assumption} \label{assumption6} 
{\bf \em The same cumulative water loss, which would have killed the organism at, or before, the time of eclosion, will kill the as yet unfed, teneral fly.} \end{assumption}
The assumption is that the total, cumulative water loss since parturition, can
be interpreted as if it were solely a pupal water loss to predict survival up
until the first meal. By using the pupal formulae alone, that water loss can be
redefined in terms of a soil humidity value at \mbox{24~$^\circ\mathrm{C}$} and,
consequently, a measure of survival as per Fig. \ref{allSpeciesTogether}; in
spite of the different teneral mode of transpiration having contributed to that
total cumulative water loss. 

The format in which the only available emergence data was
originally produced necessitates that pupal emergence be taken to be that for
the pertinent soil humidity at 24~$^\circ\mathrm{C}$, that which produced an
identical pupal water loss (Childs\nocite{Childs1}, 2013 and 2009). 
\begin{assumption} \label{assumption5} 
{\bf \em The survival for a given water loss, is the same as the pupal emergence for the steady humidity at 24$^{\mbox{o}}$C, that produced an equivalent total water loss.} 
\end{assumption}
To determine survival, water loss was re-expressed as an humidity which produced
an identical, total water loss in the pupa at 24~$^\circ\mathrm{C}$. The
percentage emergence is therefore given by some function, $E(
h_{\mbox{\scriptsize 24}}( k_{\mbox{\scriptsize pupal}} ) )$, where
$h_{\mbox{\scriptsize 24}}$ is the humidity, at 24~$^\circ\mathrm{C}$, which
gives rise to an identical water loss in the pupa, $k_{\mbox{\scriptsize
pupal}}$, to that calculated to have been incurred by the teneral under the
conditions in question. 

Lastly, the same explanation for the pupal emergence data as in
Childs\nocite{Childs1}\nocite{Childs2} (2013 and 2009) was assumed. Some pupae
will be slightly bigger, have slightly bigger reserves and more competent
integuments. Yet others will be slightly smaller, have slightly smaller reserves
and less competent integuments. This justifies the following interpretation of
the Bursell\nocite{Bursell1} (1958) and Buxton and Lewis\nocite{BuxtonAndLewis1}
(1934) pupal emergence data.
\begin{assumption} \label{assumption4} 
{\bf \em The relationship between pupal emergence and humidity, at 24~$^\circ\mathrm{C}$, is a Gaussian curve, or a part thereof. 
} 
\end{assumption}  
The parameters in 
\begin{eqnarray}
E(h_{\mbox{\scriptsize 24}}) &=& a \ \mathop{\rm exp}\left[ - \frac{(h_{\mbox{\scriptsize 24}} - b)^2}{2 c^2} \right], \nonumber
\end{eqnarray}
for each species, are provided in Table \ref{emergence}.
\begin{table}[H]
\begin{center}
\begin{tabular}{l l c c c}  
&  &  &  & \\
~ ~ group & ~ species & $a$ & $b$ & $c$ \\ 
&  &  &  & \\ \hline 
&  &  &  & \\
{\em morsitans} & {\em austeni} \ & $101.663$ & $73.1591$ & $30.6468$ \\ 
 \ & {\em morsitans} &  $94.4792$ & $70.6391$ & $77.3495$ \\ 
 \ & {\em pallidipes} & $86.6257$ & $71.5636$ & $54.9713$ \\ 
 \ & {\em submorsitans} & $94.5092$ & $81.1895$ & $75.4474$ \\ 
 \ & {\em swynnertoni} & $ 94.0194$ & $62.4064$ & $75.2339$ \\ 
&  &  &  & \\
{\em palpalis} & {\em palpalis} & $95.8732$ & $78.8419$ & $23.4835$ \\
 \ & {\em tachinoides} & $98.8383$ & $79.6877$ & $40.8616$ \\ 
&  &  &  & \\
{\em fusca} & {\em brevipalpis} \ &  $94.0057$ & $84.0199$ & $13.6433$ \\ 
&  &  &  & \\
\end{tabular}
\caption{Parameters for the fit of a Gaussian curve to the
Bursell\nocite{Bursell2} (1959) and Buxton and Lewis\nocite{BuxtonAndLewis1}
(1934) pupal emergence data for a variety of species (Childs, 20013 and 2009).
All are at \mbox{24~$^\circ\mathrm{C}$}, except {\em G. tachinoides}
(30~$^\circ\mathrm{C}$).} \label{emergence}
\end{center}
\end{table}
The assumption produced a pleasing result (Fig. \ref{allSpeciesTogether}), how
ever correct, or otherwise, the underlying reasoning. Note that no attempt has
been made to accommodate the Bursell\nocite{Bursell1} (1958) opinion that all
\mbox{Fig. \ref{allSpeciesTogether}} curves should be shifted to the left by
approximately 10\% $\mathrm{r.h.}$, due to the slightly inferior quality of the
pupal material he used. 
 \nocite{Bursell1}\nocite{Childs2}

\section{Results}

\begin{figure}[H]
    \begin{center}
\includegraphics[height=11cm, angle=0, clip = true]{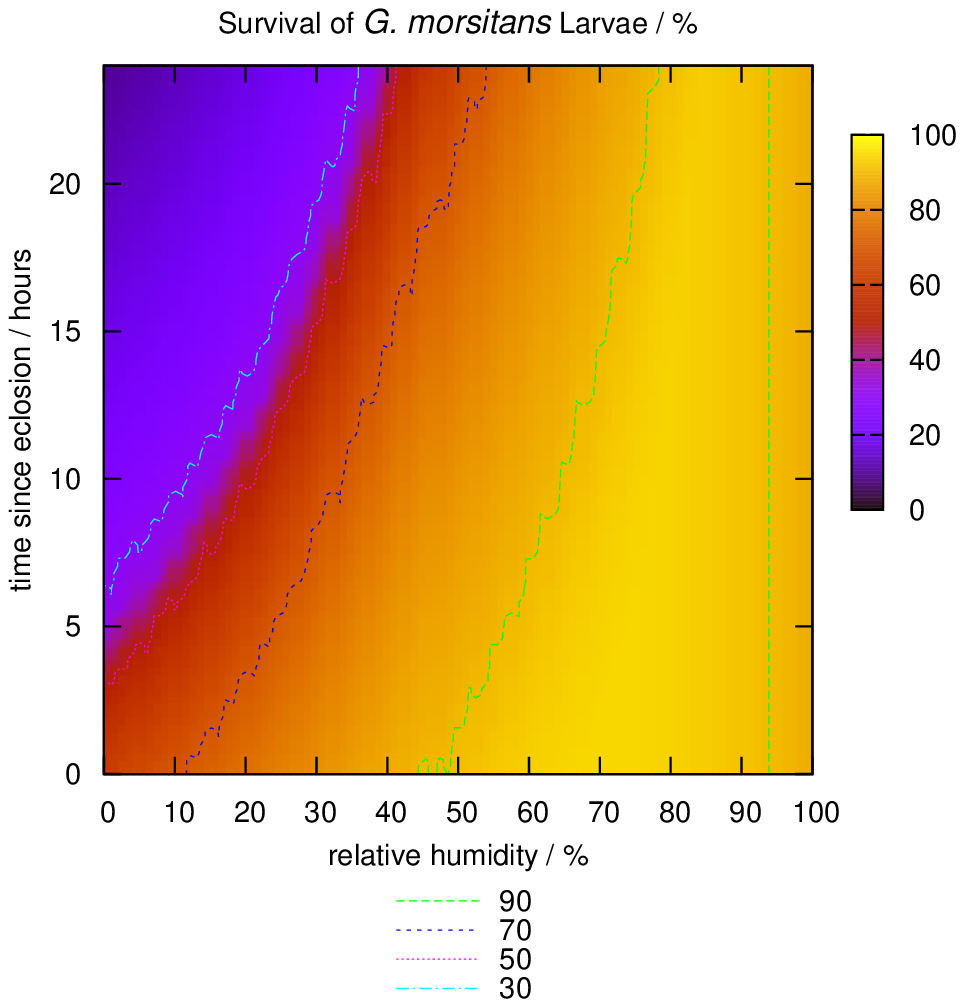}
\includegraphics[height=11cm, angle=0, clip = true]{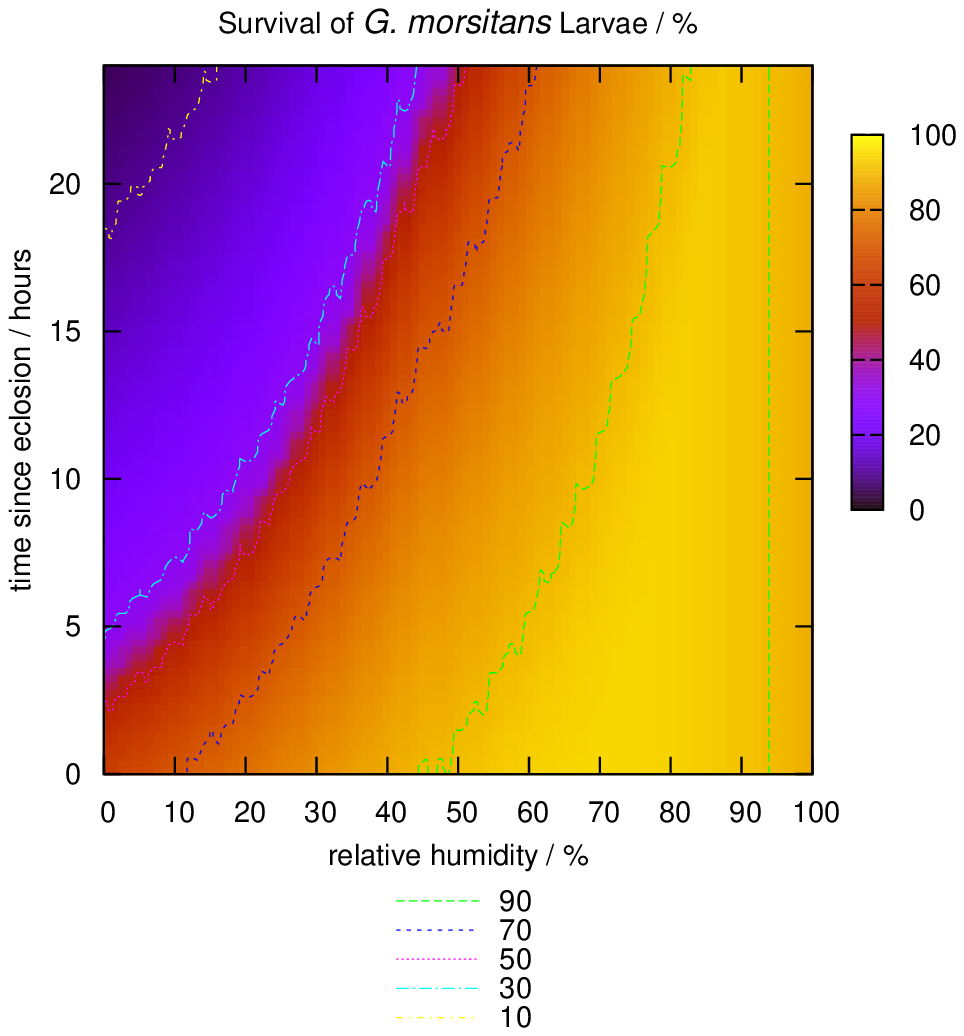}
\includegraphics[height=11cm, angle=0, clip = true]{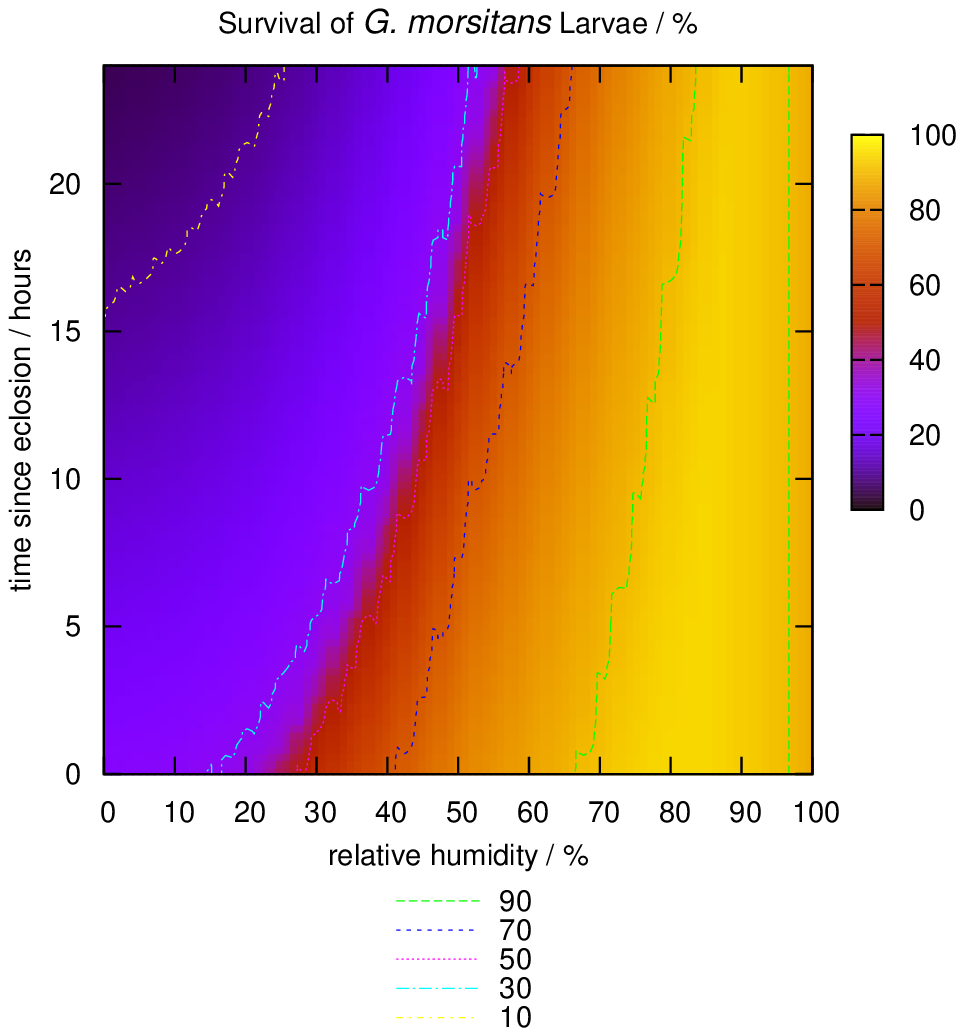}
\includegraphics[height=11cm, angle=0, clip = true]{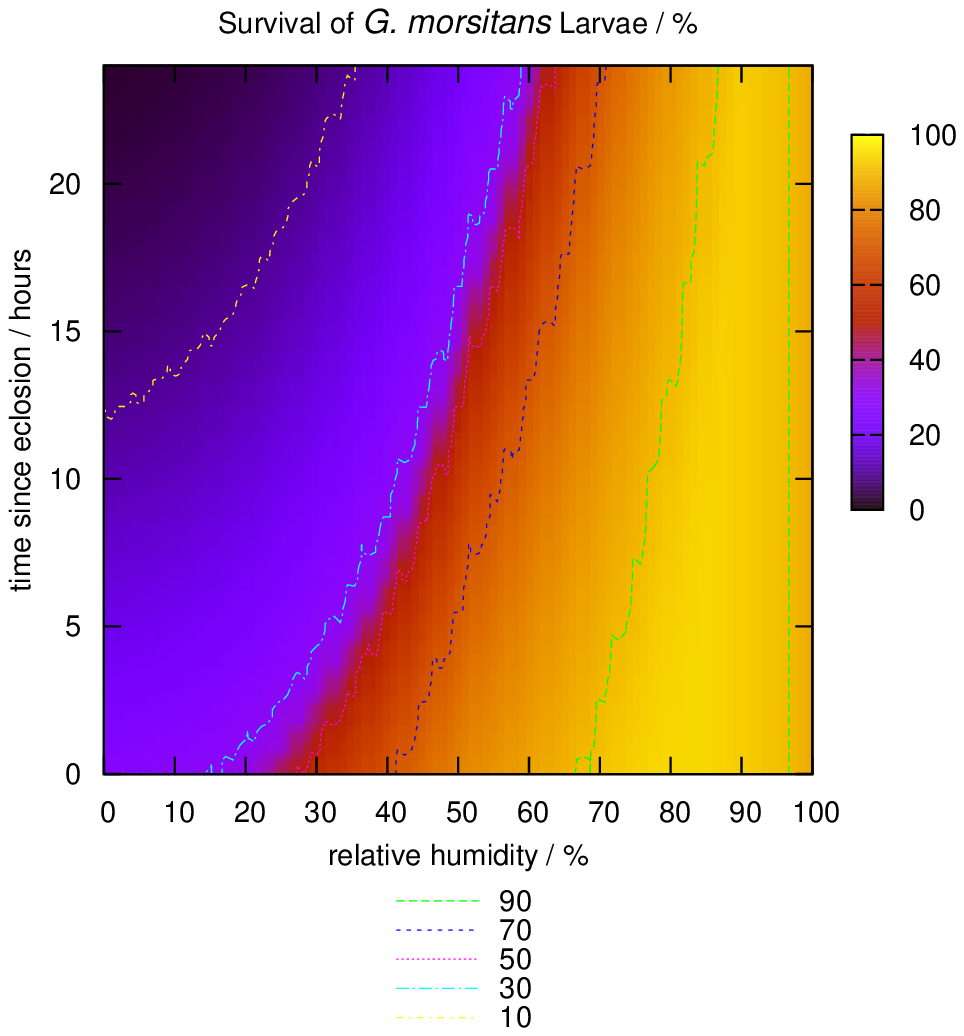}
\caption{The effect of activity on survival, since parturition, in {\em G. morsitans} tenerals: At top left, a 10\% activity level at 25~$^\circ\mathrm{C}$. At top right, a 30\% activity level at 25~$^\circ\mathrm{C}$. At bottom left, a 10\% activity level at 30~$^\circ\mathrm{C}$. At bottom right, a 30\% activity level at 30~$^\circ\mathrm{C}$.} \label{teneralSurvivalAt25degrees}
   \end{center}
\end{figure}

\begin{figure}[H]
    \begin{center}
\includegraphics[height=11cm, angle=0, clip = true]{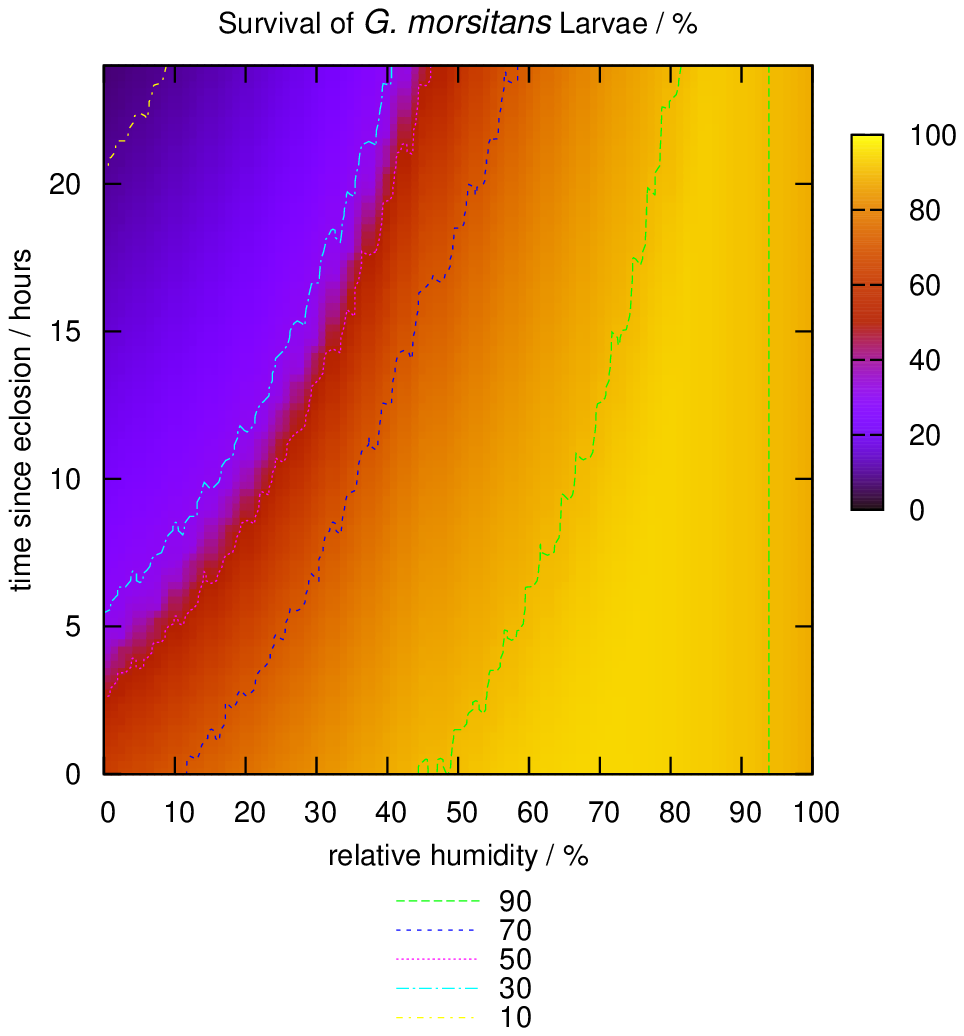}
\includegraphics[height=11cm, angle=0, clip = true]{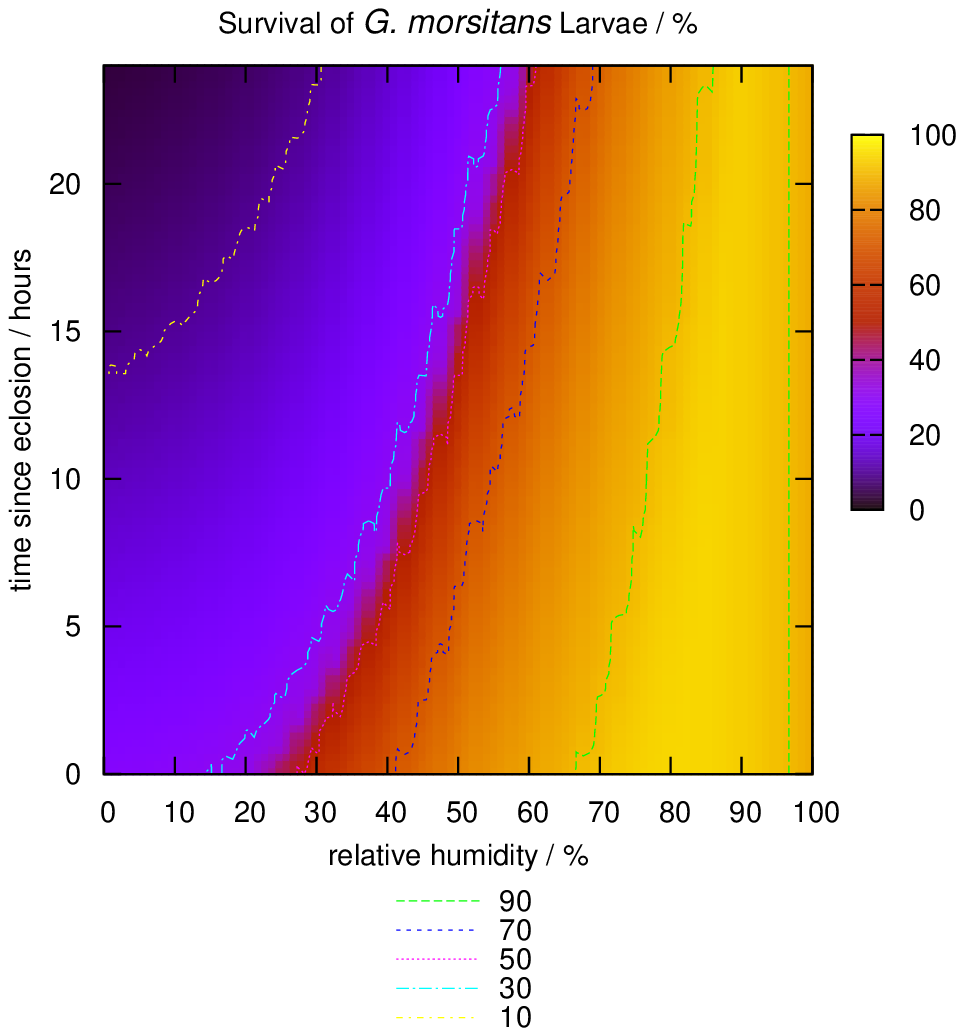}
\includegraphics[height=11cm, angle=0, clip = true]{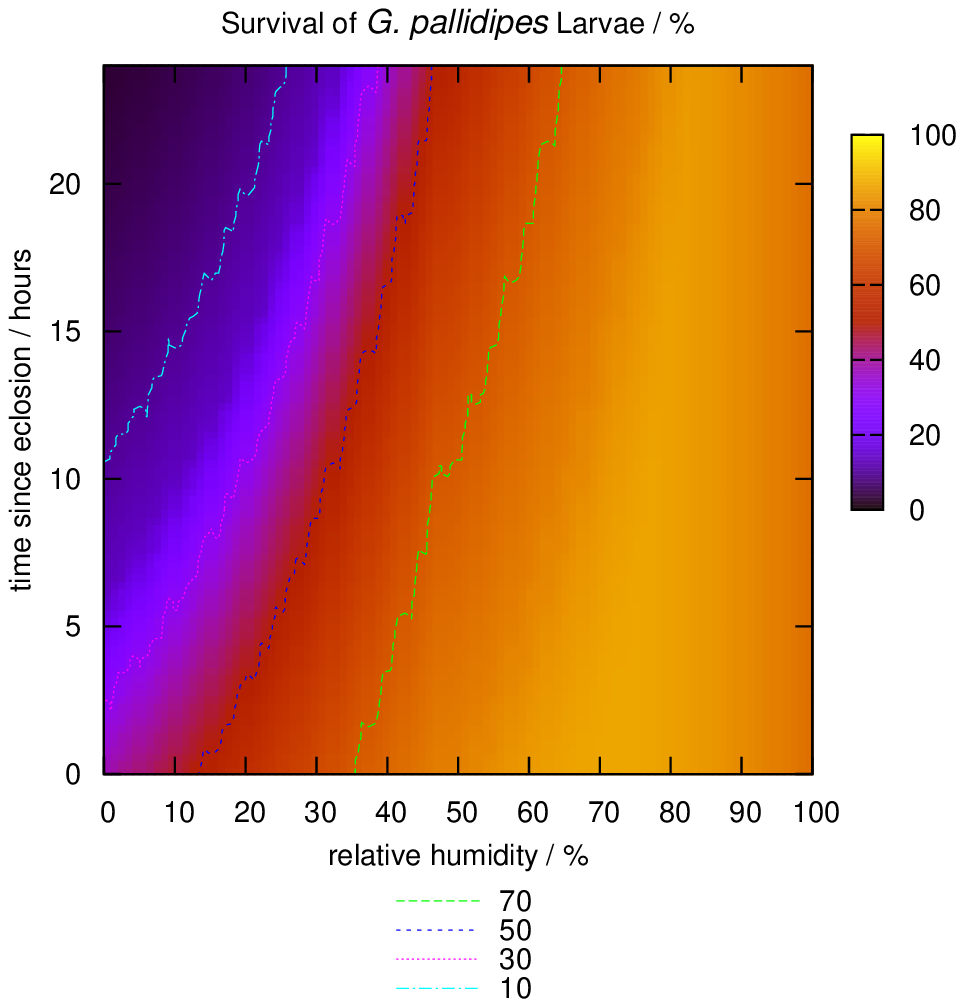}
\includegraphics[height=11cm, angle=0, clip = true]{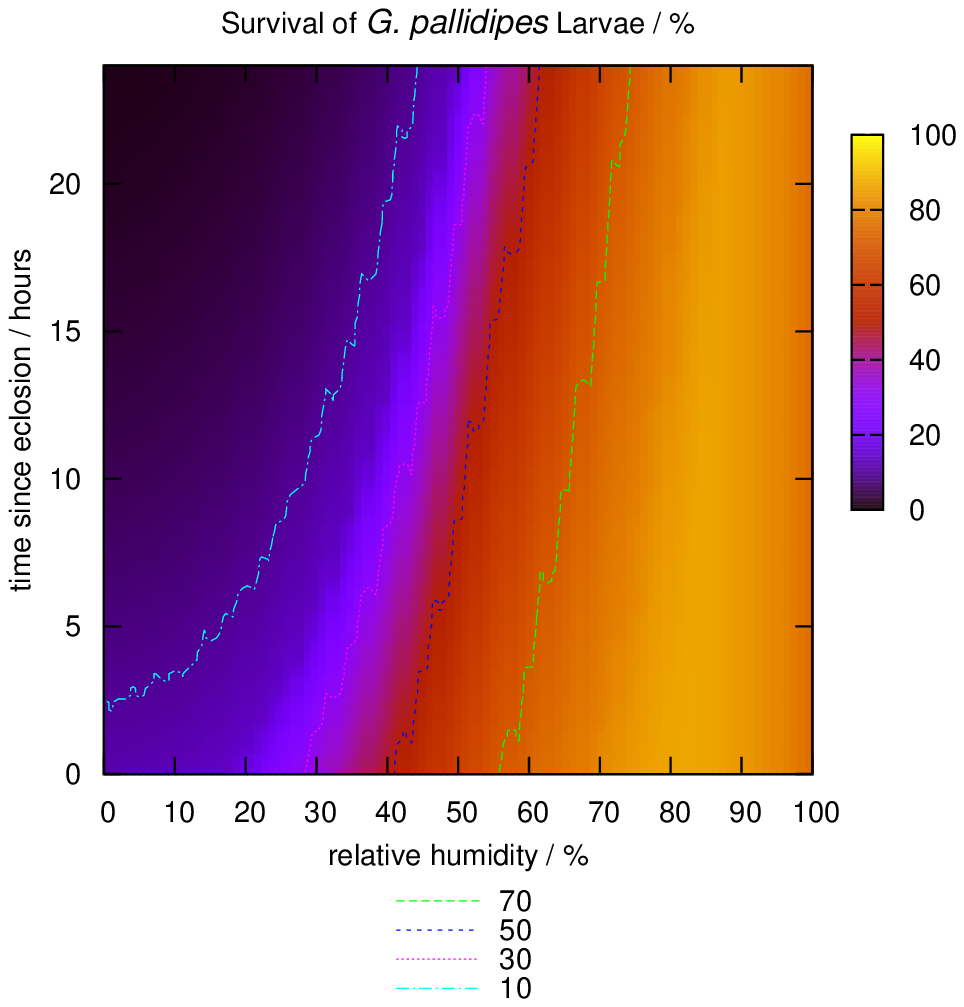}
\caption{Survival, since parturition, in tenerals at 25~$^\circ\mathrm{C}$ and a 20\% activity level. At top left, for a pupal phase at 25~$^\circ\mathrm{C}$ in {\em G. morsitans}. At top right, for a pupal phase at 30~$^\circ\mathrm{C}$ in \mbox{{\em G. morsitans}}. At bottom left, for a pupal phase at 25~$^\circ\mathrm{C}$ in {\em G. pallidipes}. At bottom right, for a pupal phase at 30~$^\circ\mathrm{C}$ in {\em G. pallidipes}.} \label{morsitansSurvival}
   \end{center}
\end{figure}

\begin{figure}[H]
    \begin{center}
\includegraphics[height=11cm, angle=0, clip = true]{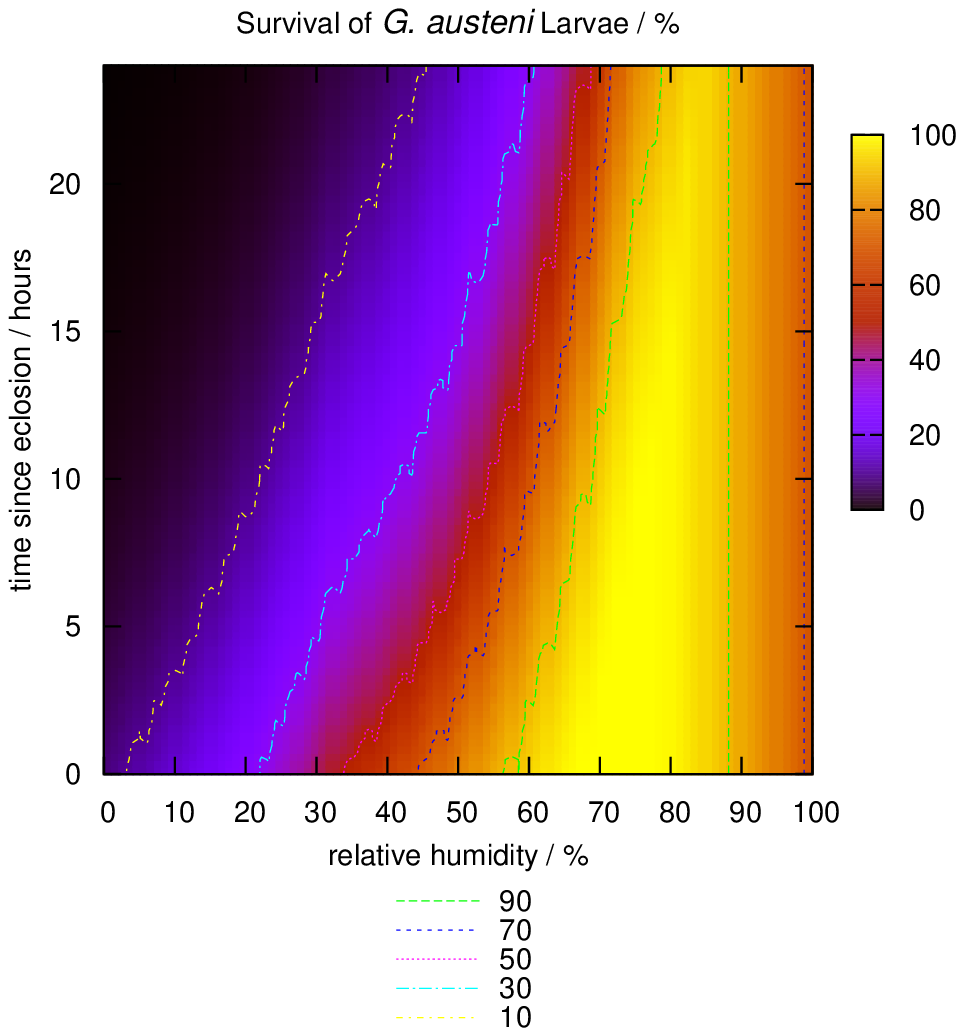}
\includegraphics[height=11cm, angle=0, clip = true]{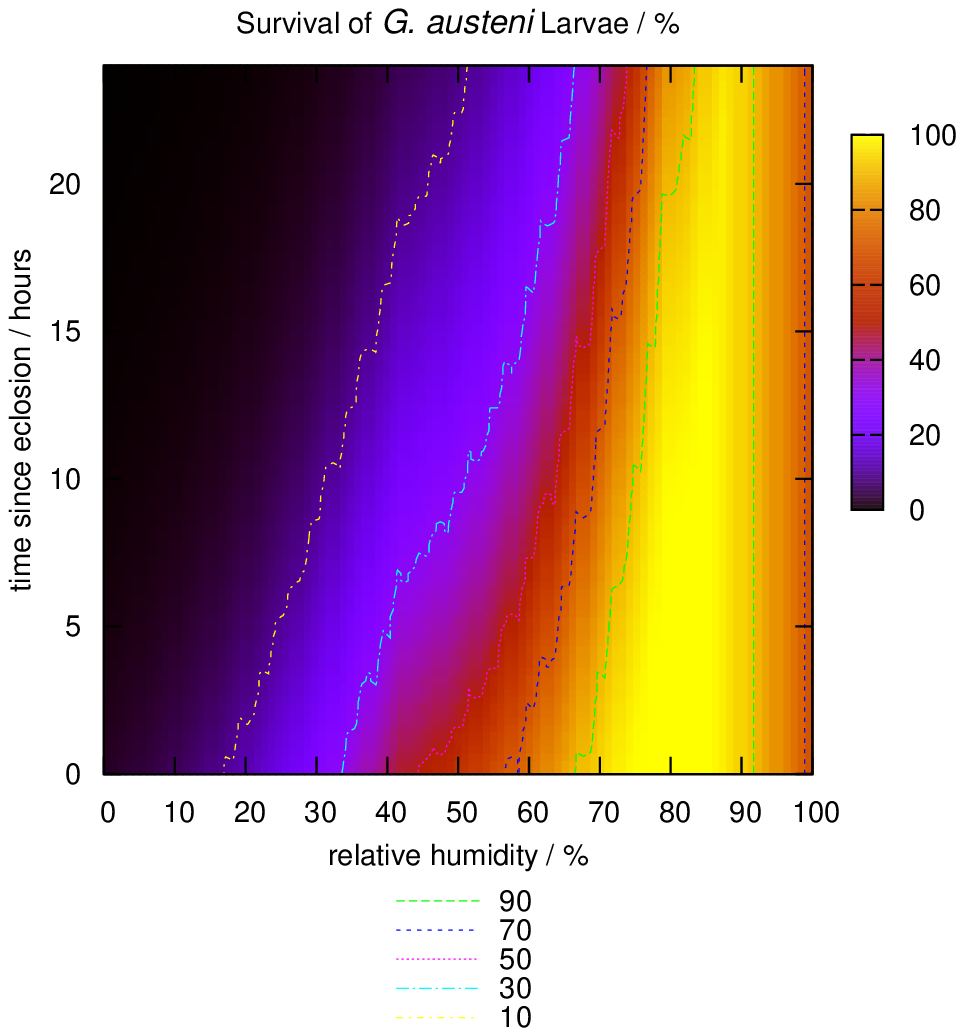}
\includegraphics[height=11cm, angle=0, clip = true]{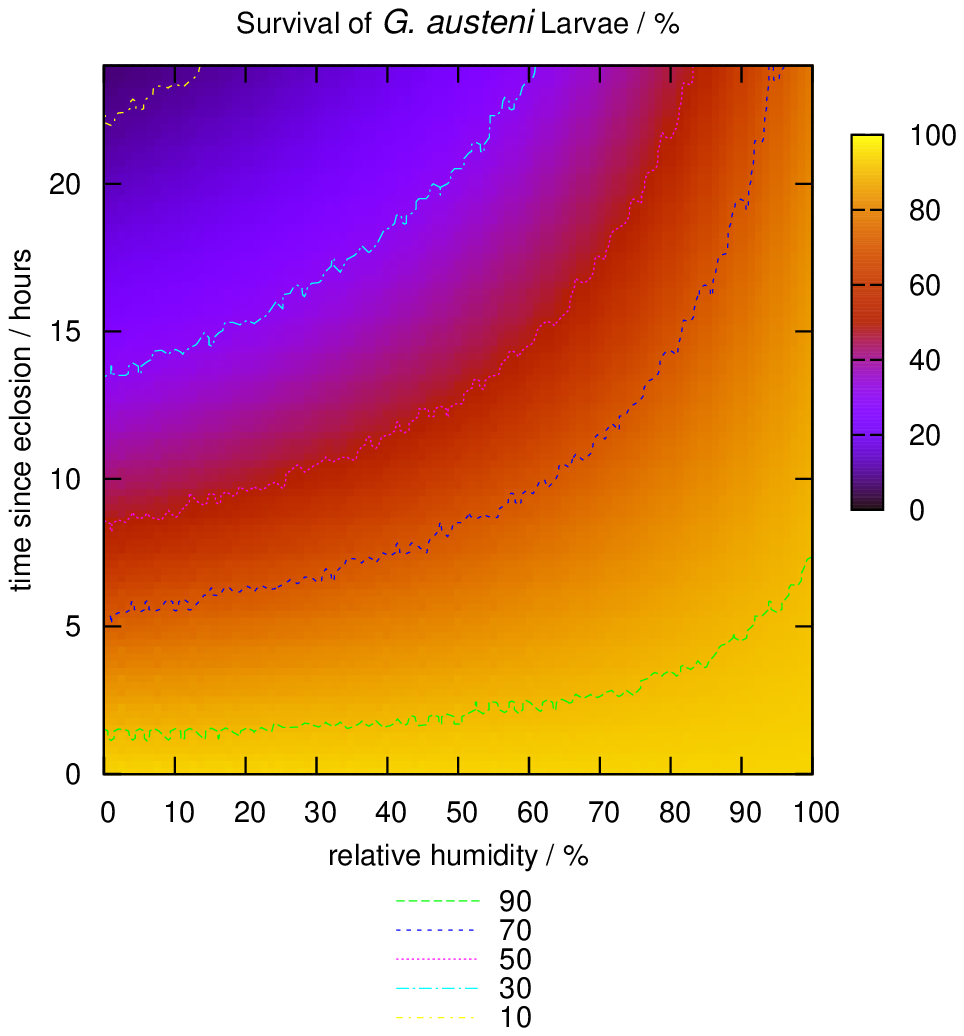}
\includegraphics[height=11cm, angle=0, clip = true]{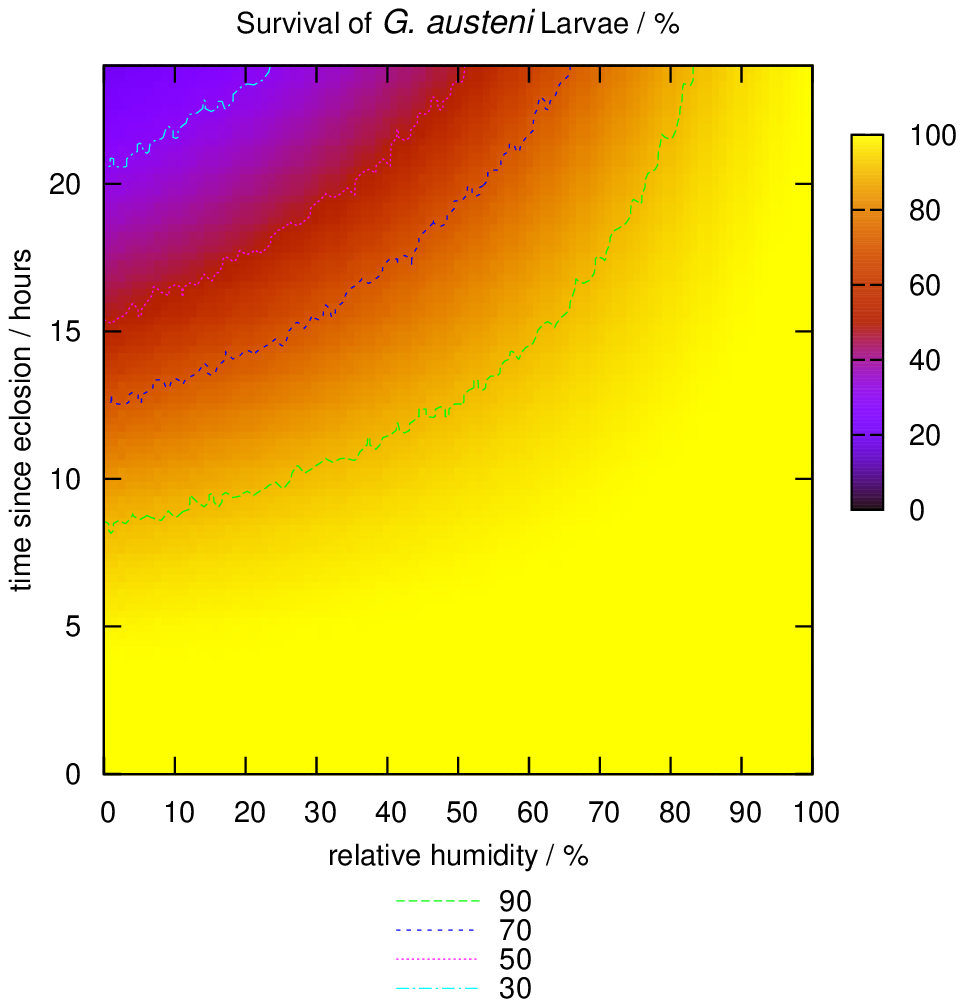}
\caption{Survival, since parturition, in {\em G. austeni} tenerals at 25
$^\circ\mathrm{C}$ and a 20\% activity level. At top left, for a pupal phase at
25~$^\circ\mathrm{C}$. At top right, for a pupal phase at 30~$^\circ\mathrm{C}$.
At bottom left, for a pupal phase at 25~$^\circ\mathrm{C}$ and 60\%
$\mathrm{r.h.}$. At bottom right, for a pupal phase at 25~$^\circ\mathrm{C}$ and
\mbox{75\% $\mathrm{r.h.}$}} \label{austeniSurvival}
   \end{center}
\end{figure}

\begin{figure}[H]
    \begin{center}
\includegraphics[height=11cm, angle=0, clip = true]{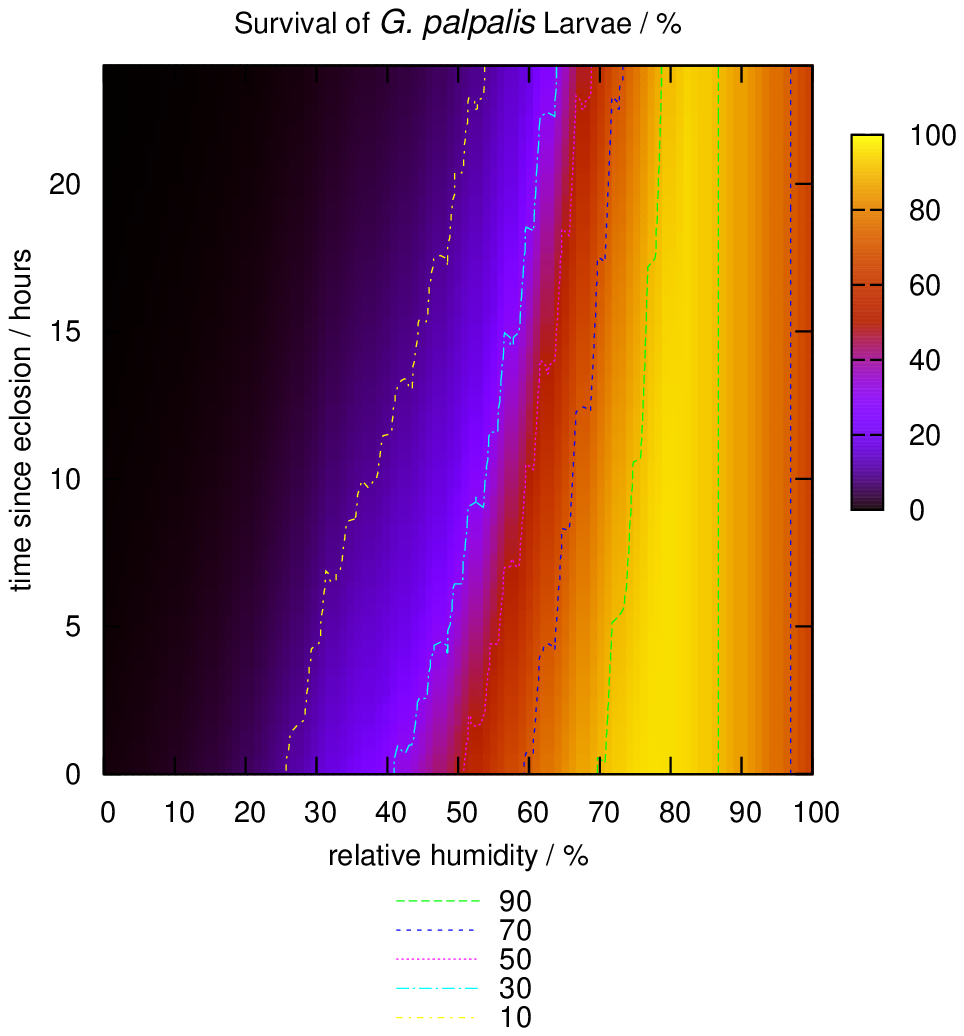}
\includegraphics[height=11cm, angle=0, clip = true]{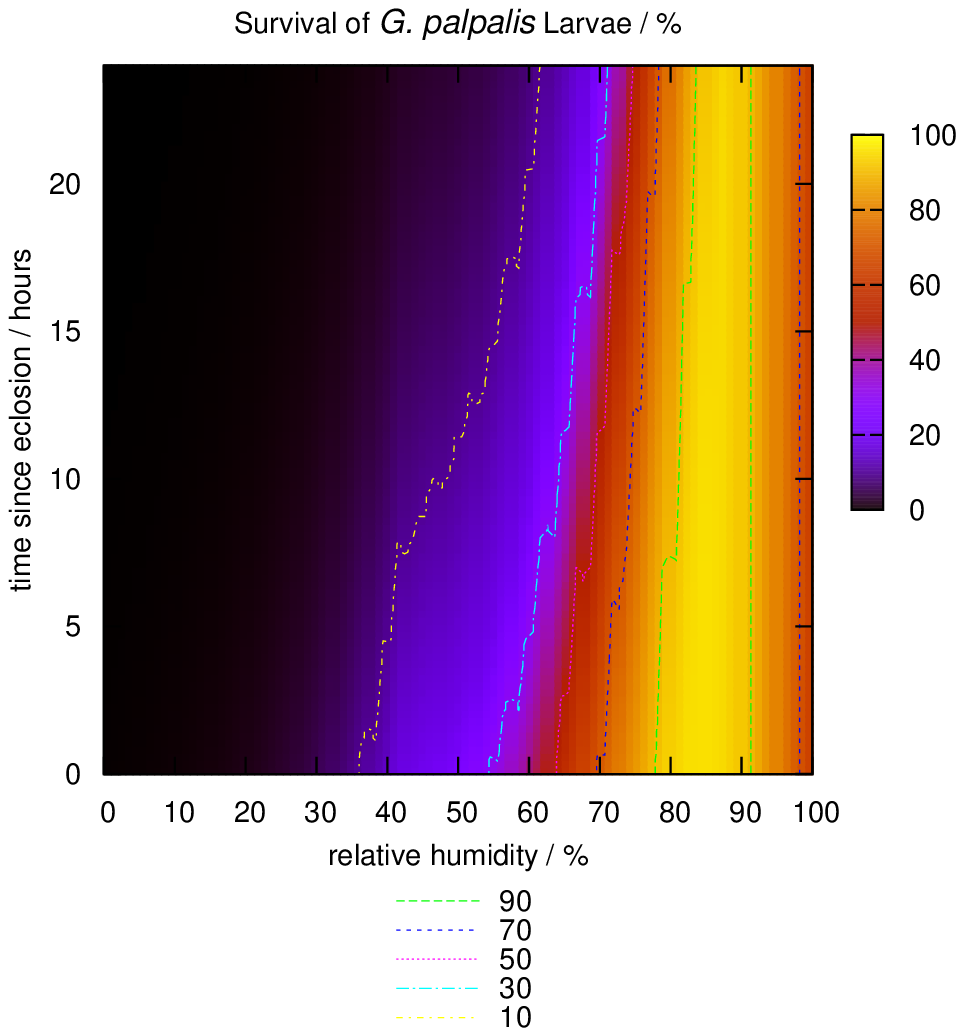}
\includegraphics[height=11cm, angle=0, clip = true]{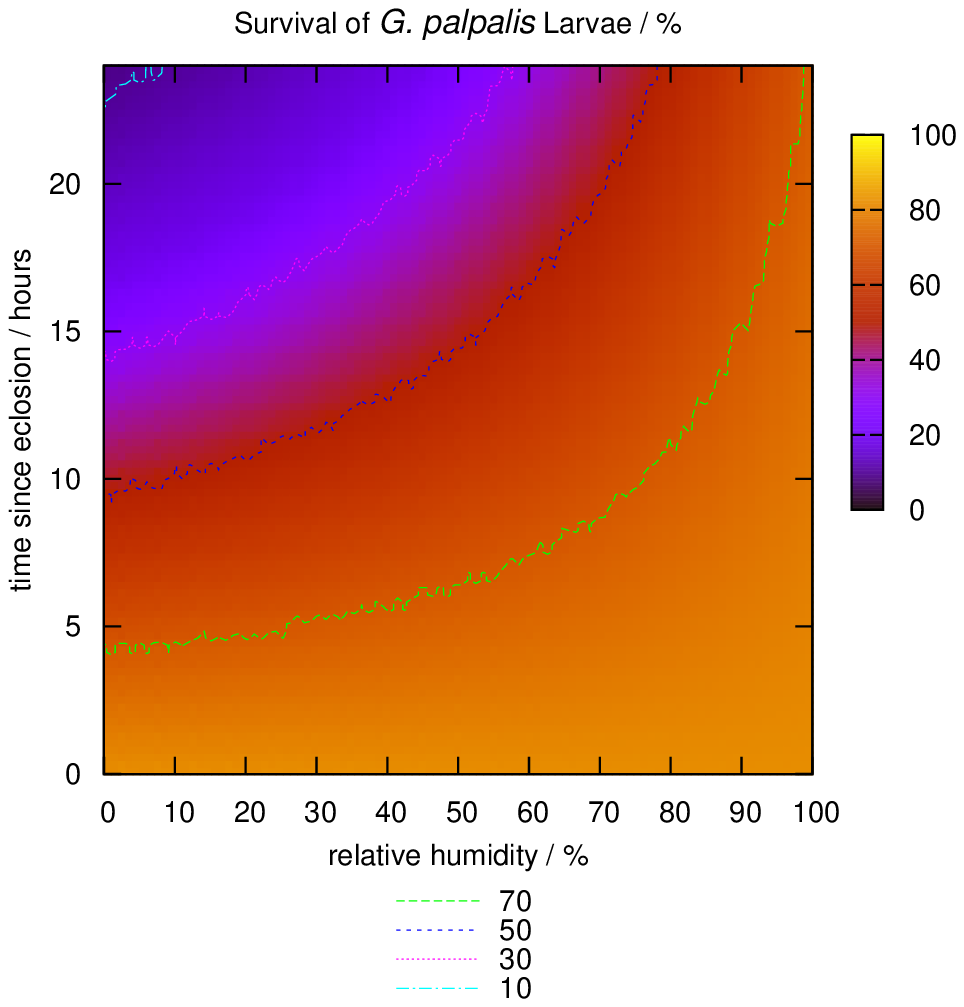}
\includegraphics[height=11cm, angle=0, clip = true]{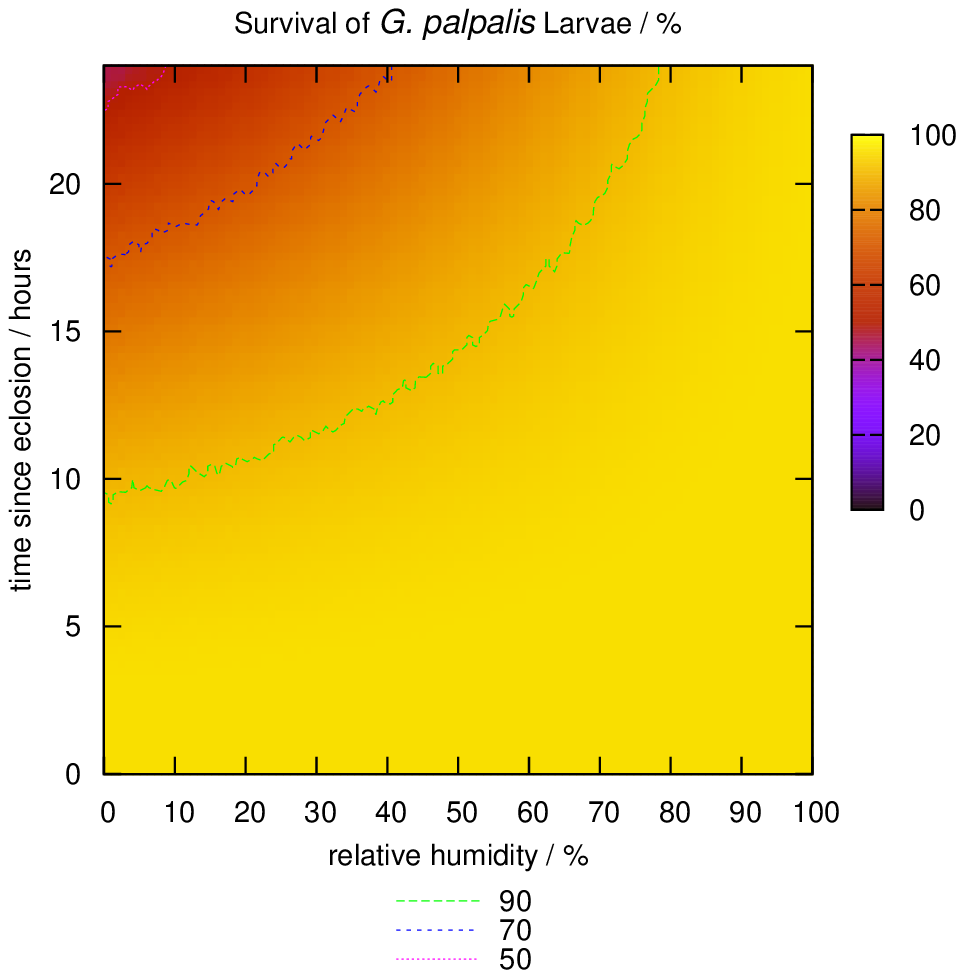}
\caption{Survival, since parturition, in {\em G. palpalis} tenerals at 25
$^\circ\mathrm{C}$ and a 20\% activity level. At top left, for a pupal phase at
25~$^\circ\mathrm{C}$. At top right, for a pupal phase at 30~$^\circ\mathrm{C}$.
At bottom left, for a pupal phase at 25~$^\circ\mathrm{C}$ and 65\%
$\mathrm{r.h.}$. At bottom right, for a pupal phase at 25~$^\circ\mathrm{C}$ and
\mbox{80\% $\mathrm{r.h.}$}} \label{palpalisSurvival}
   \end{center}
\end{figure}

\begin{figure}[H]
    \begin{center}
\includegraphics[height=11cm, angle=0, clip = true]{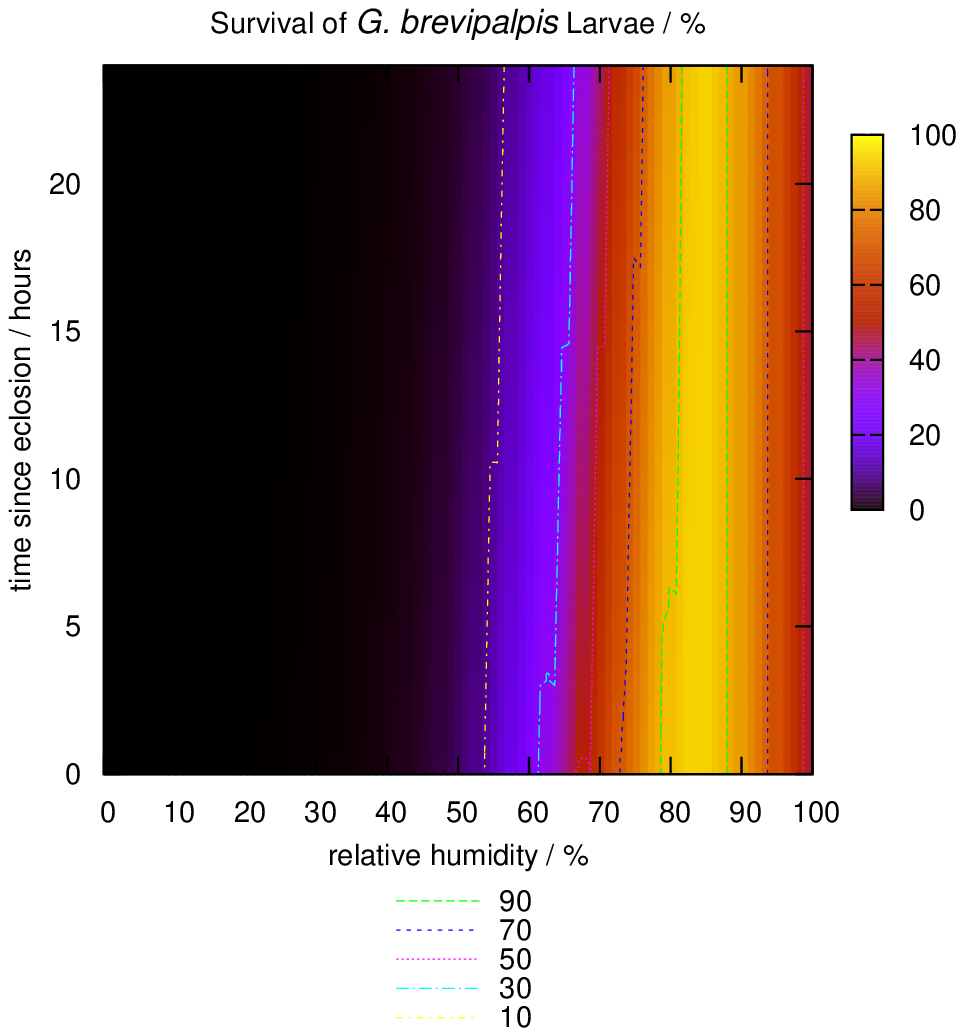}
\includegraphics[height=11cm, angle=0, clip = true]{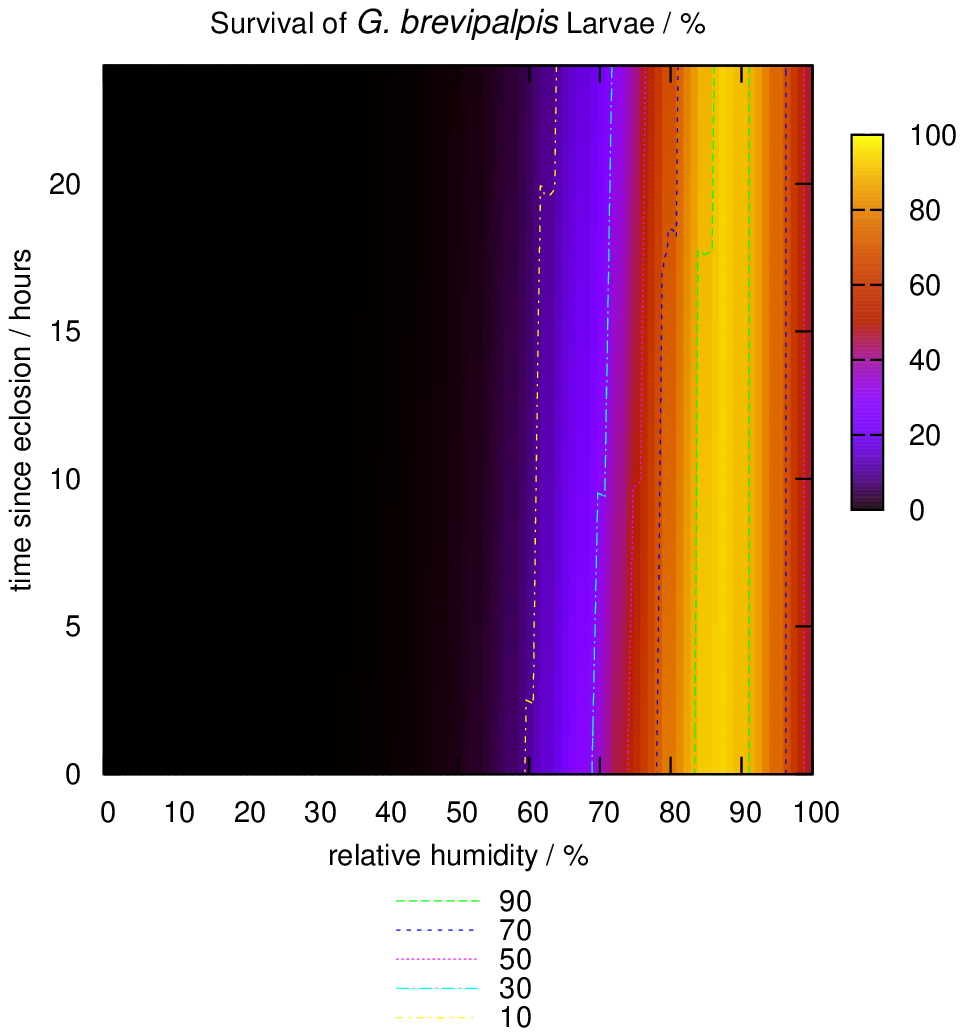}
\includegraphics[height=11cm, angle=0, clip = true]{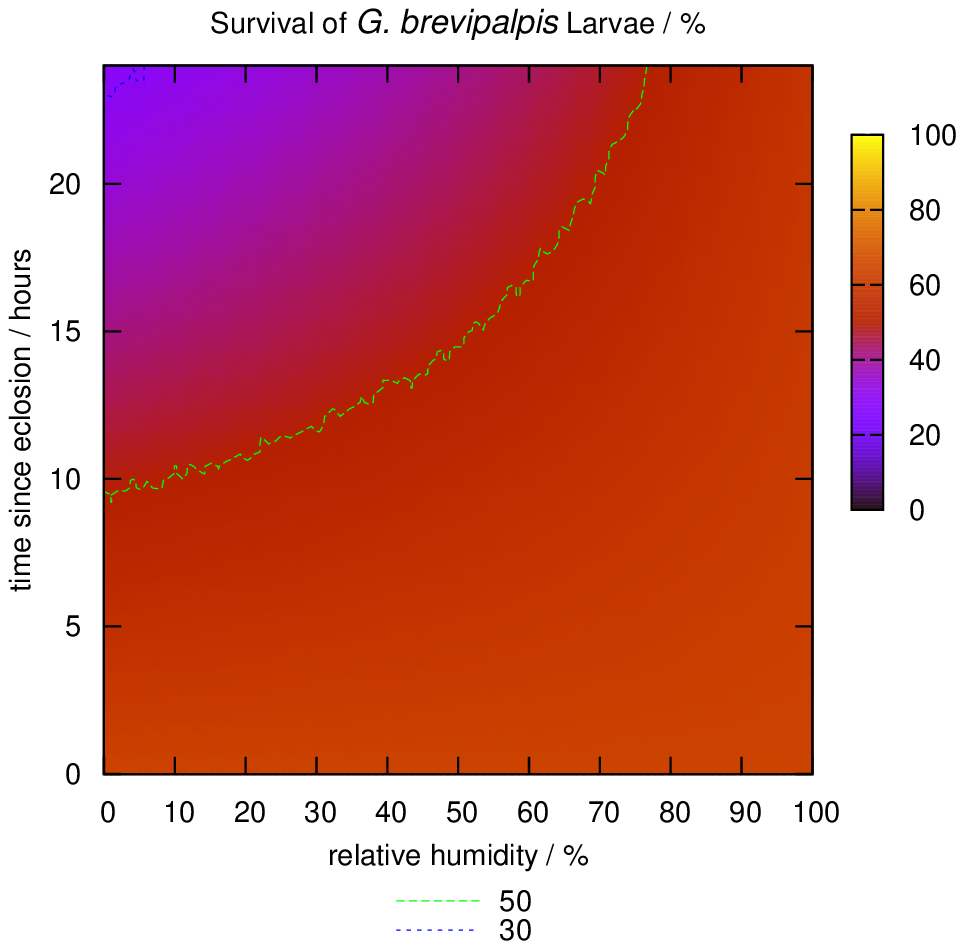}
\includegraphics[height=11cm, angle=0, clip = true]{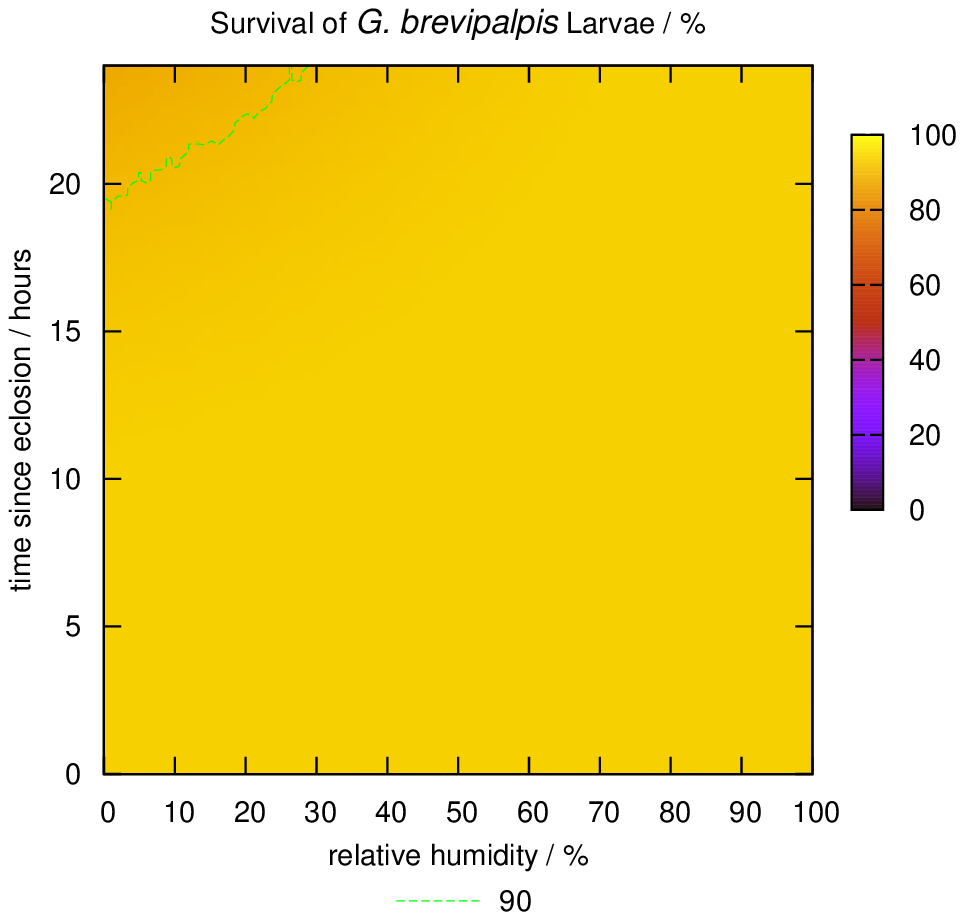}
\caption{Survival, since parturition, in {\em G. brevipalpis} tenerals at 25
$^\circ\mathrm{C}$ and a 20\% activity level. At top left, for a pupal phase at
25~$^\circ\mathrm{C}$. At top right, for a pupal phase at 30~$^\circ\mathrm{C}$.
At bottom left, for a pupal phase at 25~$^\circ\mathrm{C}$ and 71\%
$\mathrm{r.h.}$. At bottom right, for a pupal phase at 25~$^\circ\mathrm{C}$ and
85\% $\mathrm{r.h.}$} \label{brevipalpisSurvival}
   \end{center}
\end{figure}

\begin{figure}[H]
    \begin{center}
\includegraphics[height=11cm, angle=0, clip = true]{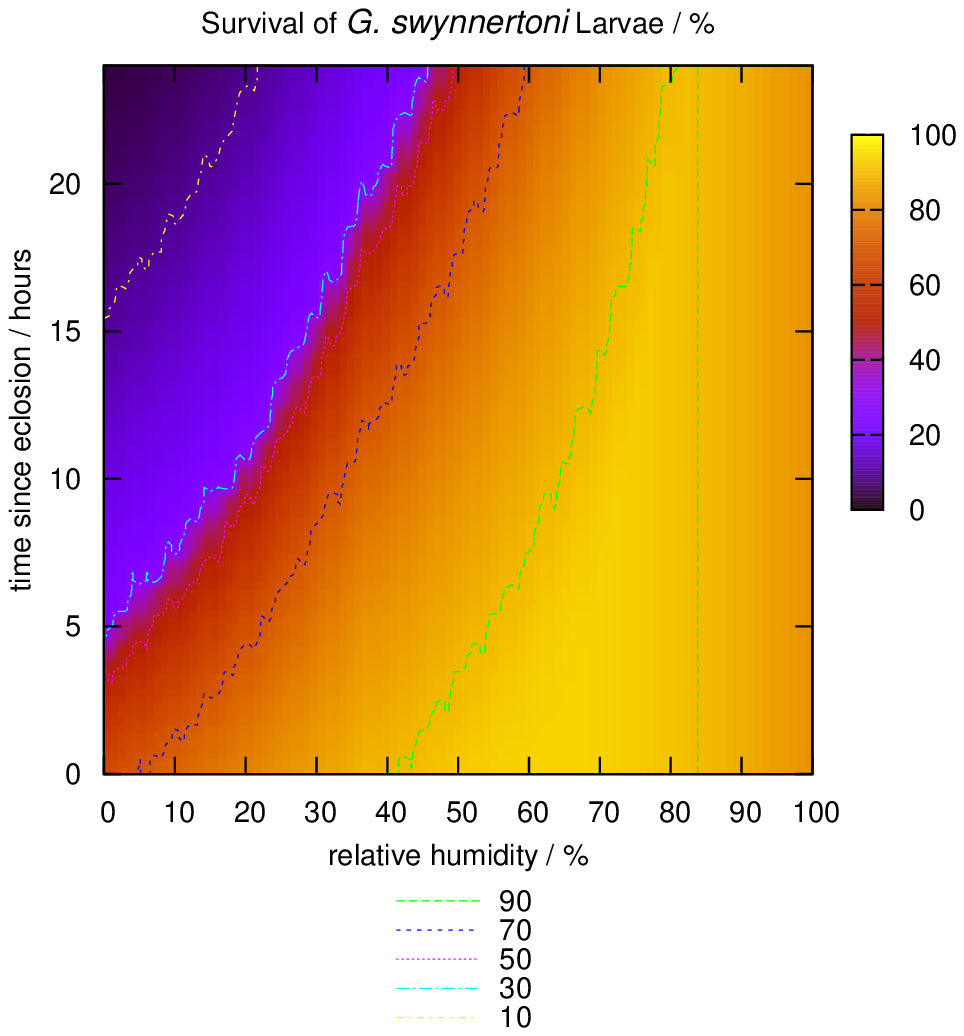}
\includegraphics[height=11cm, angle=0, clip = true]{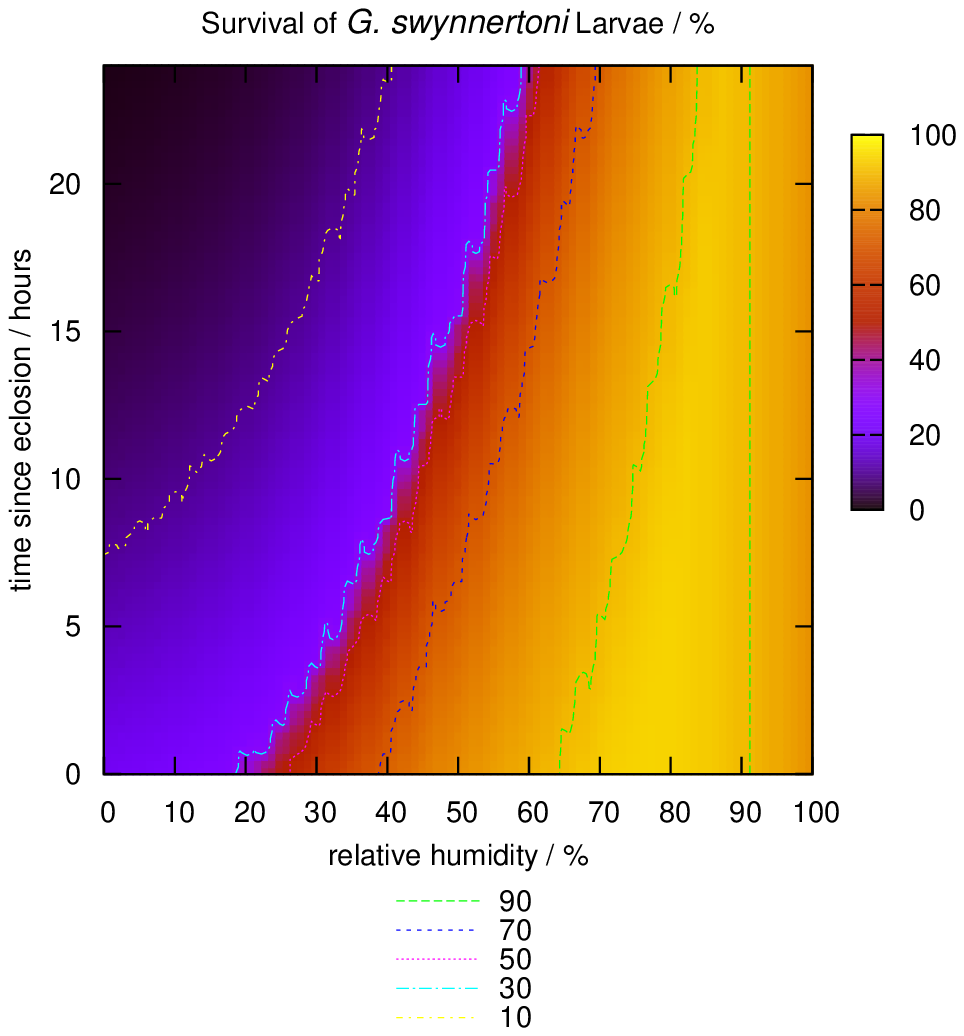}
\caption{Survival, since parturition, in tenerals at 25~$^\circ\mathrm{C}$ and a
20\% activity level. At left, for a pupal phase at 25~$^\circ\mathrm{C}$ in
{\em G. swynnertoni}. At right, for a pupal phase at 30
$^\circ\mathrm{C}$ in {\em G. swynnertoni}.} \label{swynnertoniSurvival}
   \end{center}
\end{figure}

Fig. \ref{teneralSurvivalAt25degrees} explores the effect of activity on {\em G.
morsitans} water loss. The Figs.
\ref{morsitansSurvival}--\ref{swynnertoniSurvival} composites show, firstly, the
combined pupal-teneral response to both an atmosphere and soil of identical
humidity, at 25~$^\circ\mathrm{C}$ and a 20\% activity level, for each species.
The temperature of the pupal environment is then raised to 30~$^\circ\mathrm{C}$
to produce a second plot for each species. Such scenarios are, nonetheless,
probably not very realistic, especially in the case of more specialized species.
The fate of pupae from a single pupal environment really needs to be considered
in order to examine the nature of the teneral stage in isolation. It is for this
reason that, in the case of the hygrophilic species, additional results are
plotted for pupae subjected to the drier extreme of pupal habitat from which
tenerals should survive, as well as for those from the ideal pupal environment. 

\subsection{Sources of Error}

There is no numerical error associated with the integration in the
steady-humidity and steady-activity context of these results. This is since the
time derivative is a factor in the formula for truncation error. The tolerance
in the half interval search, used to determine the level of survival, is several
decimal places. All numerical error otherwise derives from the pupal stage.
There are, doubtless, a multifarious bevy of other errors, their sources ranging
from the sparseness and errors in the data itself, to the functions fitted, the
formula for the puparial duration and, very likely, Assumptions
\ref{assumption1} and \ref{assumption3} in particular. The standard error in the
Bursell\nocite{Bursell2} (1959) estimates is around 10\%, there is a 2\% error
incurred in equating weight loss to water loss and there is the discrepancy in
the {\em G brevipalpis} data already mentioned. The above results are also
rooted in the unmodified, Bursell\nocite{Bursell1} (1958) data. No attempt has
been made to accommodate the opinion that the pupae were of a slightly inferior
quality and that all \mbox{Fig. \ref{allSpeciesTogether}} curves should
consequently be shifted to the left by approximately 10\% $\mathrm{r.h.}$ 

There are many other sources of error. It should, however, be pointed out that
some errors compensate for others and there are many reasons to believe that
the overall error in the results does not exceed the projected target of
15--20\%. There is reason to believe that in many instances the error incurred
is far less than this.

\section{A Probabilistic Formulation of Adult Recruitment}

While the survival level at the average time to feed could be interpreted as the
number of original larvae which survive the teneral phase to adulthood, the
following is a more rigorous method based, instead, on the probability of the
teneral feeding. I am indebted to \mbox{Dr. Neil Muller} for assisting me with
statistical aspects of what follows.

\subsection{The Probability of Not Feeding and Surviving}

Suppose one defines a value
\begin{eqnarray*} 
E_{0} &=& E( \ \mathop{\rm min}\{ b, h_{\mbox{\scriptsize 24}}(k_{\mbox{\scriptsize pupal}}) \} \ ),
\end{eqnarray*} 
in which $b$ is the mean for $E(h_{\mbox{\scriptsize 24}})$, listed in Table \ref{emergence}. Thus, the cumulative density function for the tenerals, were they never to feed and thereby become adults, would be
\begin{eqnarray} \label{111}
\frac{ E( h_{\mbox{\scriptsize 24}}( k(t) + k_{\mbox{\scriptsize pupal}} ) ) }{ E_{0} } \hspace{10mm} \mbox{for} \hspace{10mm} h_{\mbox{\scriptsize 24}} \le b, \nonumber 
\end{eqnarray} 
where the function $k(t)$ is the cumulative water loss since eclosion. Since a certain number of tenerals do manage to feed and in so doing become surviving adults, this cumulative density function is of no use as it stands. A probability density function, $\epsilon$, defined by
\begin{eqnarray} \label{2}
\epsilon(t) &=& c \frac{1}{ E_{0} } \frac{dE}{dt} \nonumber
\end{eqnarray} 
is, instead, what is required, $c$ being a normalisation constant. Since
\begin{eqnarray} \label{222}
c \int_{t_0}^{\infty} \epsilon(t) dt &=& c \left( 0 - 1 \right) \ \ = \ \ 1 \nonumber \\ 
\Rightarrow \ \epsilon(t) &=& \frac{1}{ E_{0} } \frac{ a (h_{\mbox{\scriptsize 24}} - b) }{c^2} \ e^{ - (h_{\mbox{\scriptsize 24}} - b)^2 / 2 c^2} \ \frac{ d h_{\mbox{\scriptsize 24}} }{dk} \frac{dk}{dt}, \nonumber 
\end{eqnarray} 
in which $t_0$ is the time associated with the water loss, in turn associated
with the humidity defining $E_0$ (if $t_0 \ne 0$ then the integrand is defined
as $0$ on $[0, t_0)$). This is the probability of a single fly not dying due to
dehydration over some time interval. 

If one devises a similar probability density function for flies not feeding, say $p(t)$, then the probability of a teneral still being a teneral, over some interval in time, is the number that neither fed, nor died over that time interval. If one defines
\begin{eqnarray} \label{3}
f(t) &=& \left\{ \begin{array}{l} \ p(t) \\ \displaystyle \frac{1}{ E_{0} } \frac{ a (h_{\mbox{\scriptsize 24}} - b) }{c^2} \ e^{ - (h_{\mbox{\scriptsize 24}} - b)^2 / 2 c^2} \left( \frac{dk}{ d h_{\mbox{\scriptsize 24}} } \right)^{-1} \frac{dk}{dt} \ p(t) \end{array} \right. 
\hspace{5mm} \mbox{for} \hspace{5mm} 
\begin{array}{l} h_{\mbox{\scriptsize 24}} > b \\ \\ h_{\mbox{\scriptsize 24}}
\le b \end{array} \hspace{2mm} , \nonumber 
\end{eqnarray} 
where $\frac{dk}{dt}$ is given by Eq. \ref{17} and $\frac{dk}{ d h_{\mbox{\scriptsize 24}} }$ is the rate of change of pupal water loss with humidity at 24~$^\circ\mathrm{C}$, then
\begin{eqnarray} \label{4}
\mbox{tenerals}(t) &=& \int_{0}^{t} f(t') \times E(h_{\mbox{\scriptsize 24}}(k_{\mbox{\scriptsize pupal}})) \ dt', \nonumber 
\end{eqnarray} 
in which ``tenerals'' is used to denote the number of tenerals. Note that numerical values of $k$ need to be calculated for various values of $h_{\mbox{\scriptsize 24}}$ and a curve fitted so that a $\frac{dk}{ d h_{\mbox{\scriptsize 24}} }$ relation for each species might be determined.

\subsection{Adult Recruitment}

This is the probability of those that have survived without feeding, now feeding. If $p(t)$ is the probability density function for not feeding, then $1- p(t)$ is the probability density function for feeding and therefore adulthood. The number of adults which result is 
\begin{eqnarray} \label{5}
\mbox{adults}(t) &=& \int_0^{t} ( 1- p(t') ) \times \mbox{tenerals}(t') \ dt' \nonumber \\
&=& \int_0^{t} ( 1- p(t') ) \left[ \int_0^{t'} f(t'') \ \times \ E(h_{\mbox{\scriptsize 24}}(k_{\mbox{\scriptsize pupal}})) \ dt'' \frac{}{} \right] \ dt'. \nonumber 
\end{eqnarray} 
The final tally of adults is consequently
\begin{eqnarray} \label{6}
\mbox{total adults} &=& \int_0^{\infty} ( 1- p(t) ) \left[ \int_0^{t} f(t') \ \times \ E(h_{\mbox{\scriptsize 24}}(k_{\mbox{\scriptsize pupal}})) \ dt' \frac{}{} \right] \ dt. \nonumber 
\end{eqnarray} 
  
\section{Conclusions}

High water loss rates are a consequence of high levels of teneral activity in
dry air and, it is assumed, high temperature. Such conditions lead to a
dehydration of the teneral fly which can be fatal. By knowing the average time
to the first blood-meal, the percentage of original larvae which survive to
become adults can be estimated. Some pupae eclode in the late afternoon and feed
at sunset, however, most are thought to eclode in the evening (Vale et.
al.\nocite{ValeHargroveJordanLangleyAndMews} 1976) and, considering that the
exoskeleton needs a few hours to harden first, one might surmise that the
newly-eclosed, teneral fly usually has to wait through the night, until dawn, to
feed (nocturnal species such as {\em G. longipalpis} being the exception,
according to Parker\nocite{Parker2}, 2009). In Figs.
\ref{morsitansSurvival}--\ref{swynnertoniSurvival}, the angle between the
contours and the time-axis is a measure of the susceptibility to further
dehydration during the teneral phase. Surprisingly, the postulated race against
time to replenish fluid reserves appears to apply rather more to the mesophilic
and xerophilic species, in that increasing order. Very little evidence of the
same phenomenon exists in the case of hygrophilic species for which the
atmosphere and soil have an identical humidity. Only for {\em G. austeni}
\mbox{(Fig. \ref{austeniSurvival})} is there some evidence of a motive for the
teneral to expedite under the aforementioned circumstances. 

On such a basis, it is tempting to conclude that the classification of species
as hygrophilic, mesophilic and xerophilic is irrelevant. This is certainly not
the case. From the different pupal responses to an hypothetical heat wave
(Childs\nocite{Childs1}, 2013), it has already been suggested that there is no
continuous progression from {\em G. pallidipes} to {\em G. austeni}. {\em G.
austeni} appears not to simply be a more extreme adaption of {\em G.
pallidipes}. In addition to the clear limitations water loss sets for the pupae
of each species, Rogers and Robinson\nocite{RogersAndRobinson} (2004) found that
cold cloud duration (rainfall) was far and away the most frequently occurring
variable in their top five for determining the distribution of both the {\em
fusca} and {\em palpalis} groups using satellite imagery. Normalized difference
vegetation index (NDVI) ranked second by a significant margin in those two
groups and only just beat cold cloud duration for the {\em morsitans} group. It
is not too great a stretch of the imagination to entertain the possibility that
cold cloud duration and NDVI translate directly into humidity, as might
elevation in the context of low-lying, coastal deltas, through which rivers
meander, often terminating in estuaries. Rogers and
Robinson\nocite{RogersAndRobinson} (2004) also found that rainfall was even more
relevant when it came to abundance, as opposed to distribution. From these and
other facts, it is clear that the classification of species as hygrophilic,
mesophilic and xerophilic is certainly not irrelevant. As to what it is that
really defines a species as hygrophilic, a knowledge of {\em G. brevipalpis}'
environment, then experimenting with the model provides a clue.

\begin{figure}[H]
    \begin{center}
\includegraphics[height=11cm, angle=0, clip = true]{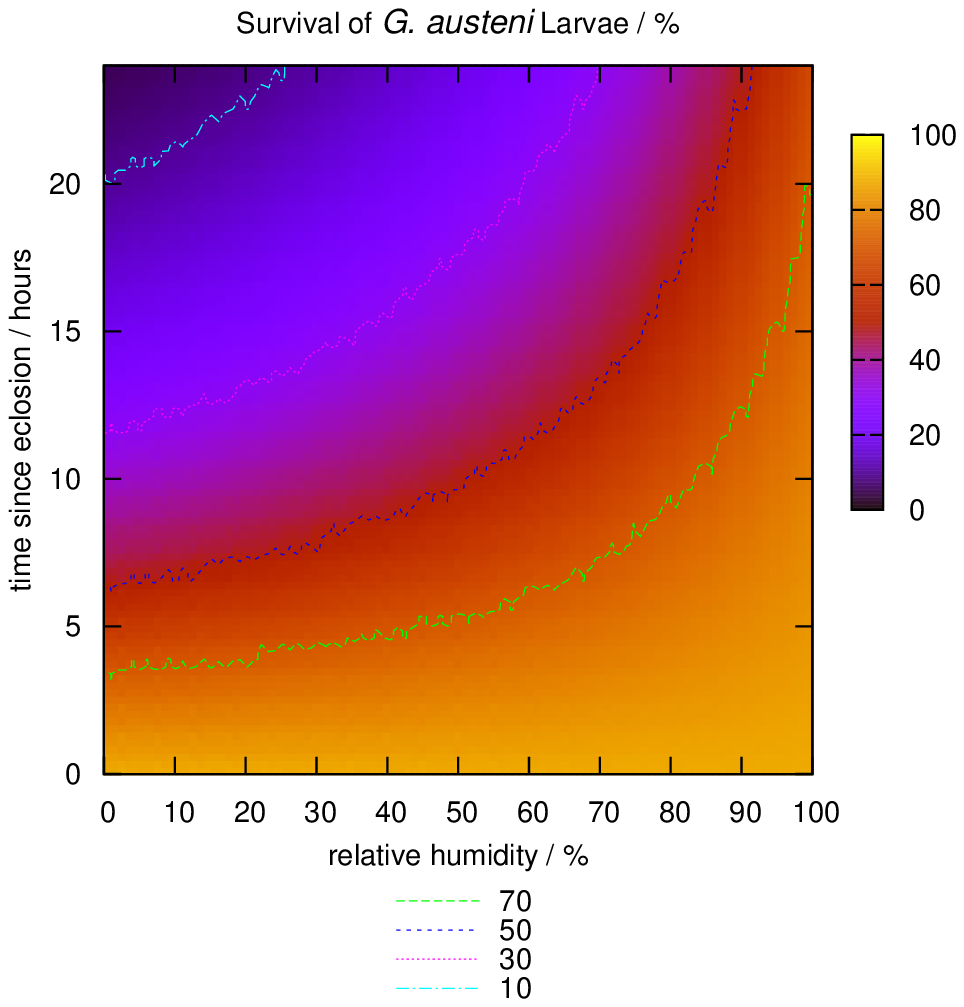}
\includegraphics[height=11cm, angle=0, clip = true]{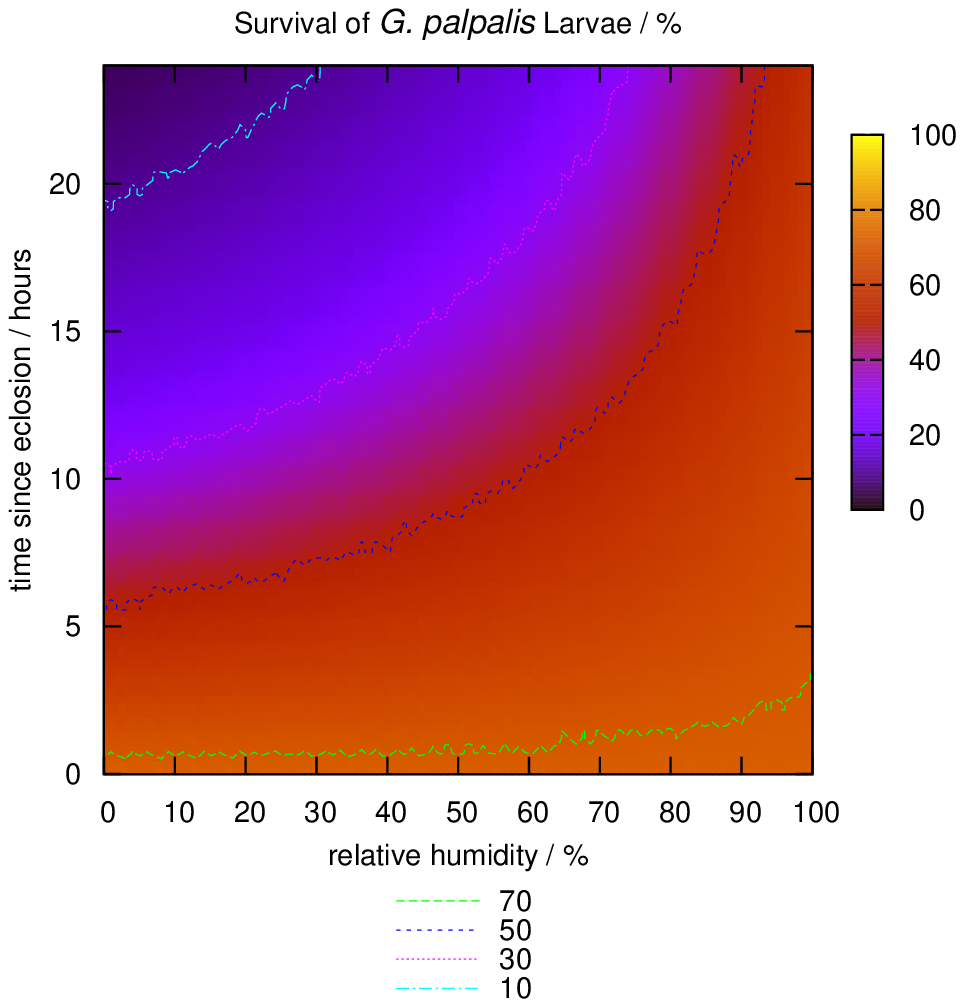}
\caption{Survival, since parturition, at 25~$^\circ\mathrm{C}$ and a 20\%
activity level. At left, in {\em G. austeni} tenerals, for a pupal phase at 55\%
$\mathrm{r.h.}$. At right, in {\em G. palpalis} tenerals, for a pupal phase at
60\% $\mathrm{r.h.}$} \label{bursellsSaturatedSoil}
   \end{center}
\end{figure}

To be fair in the case of the hygrophilic species, one really needs to consider
the fate of pupae from a single pupal environment in order to examine the nature
of the teneral stage in isolation. It is only in this context that the teneral
phase can be considered free of the fatalities suffered during the pupal stage.
The 50\% survival level might loosely be interpreted as demarcating the survival
of a single fly. More than half the {\em G. austeni} pupae still eclode for
pupal environments as dry as 40\% $\mathrm{r.h.}$ (Childs\nocite{Childs1},
2013). Despite this, only those from pupal environments as moist as 60\%
$\mathrm{r.h.}$ are resilient enough for general survival through the night,
until dawn \mbox{(Fig. \ref{austeniSurvival})}. For those that feed before
sunset and those that eclode into a very humid atmosphere, even drier pupal
substrates than this might be tolerable (Fig. \ref{bursellsSaturatedSoil}).
Those {\em G. austeni} pupae from \mbox{40--60\% $\mathrm{r.h.}$} pupal
substrates are in jeopardy for atmospheric conditions that are anything less
than ideal. For ideal pupal substrates of around 75\% $\mathrm{r.h.}$ there is
no chance of the newly-eclosed, \mbox{\em G. austeni} teneral succumbing to
dehydration (Fig. \ref{austeniSurvival}) at room temperature. The postulated
race against time to replenish fluid reserves does not exist for {\em G.
austeni}, under those circumstances. This value is validated by the fact that
Onderstepoort Veterinary Institute (O.V.I.) keep their \mbox{\em G. austeni}
colony at \mbox{75\% $\mathrm{r.h.}$} (De Beer\nocite{Chantel}, 2013). That
their specimens experience problems below 60\% $\mathrm{r.h.}$ is probably too
much of a short term effect to claim as validation, one which has more to do
with adult flies and the regularity with which they are fed. {\em G. palpalis}
is considered hygrophilic to an even greater degree than {\em G. austeni}. More
than half the {\em G. palpalis} pupae still eclode for environments as dry as
50\% $\mathrm{r.h.}$ (Childs\nocite{Childs1}, 2013). Despite this, only those
from environments as moist as 65\% $\mathrm{r.h.}$ are resilient enough for
general survival through the night, until dawn \mbox{(Fig.
\ref{palpalisSurvival})}. For those that feed before sunset and those that
eclode into a very humid atmosphere, drier pupal substrates might be tolerable
\mbox{(Fig. \ref{bursellsSaturatedSoil})}. Those {\em G. palpalis} pupae from
50--65\% $\mathrm{r.h.}$ pupal substrates are in jeopardy for atmospheric
conditions that are anything less than ideal. For ideal pupal substrates of
around 80\% $\mathrm{r.h.}$ there is no chance of the newly-eclosed, {\em G.
palpalis} teneral succumbing to dehydration (Fig. \ref{palpalisSurvival}) at
room temperature. The postulated race against time to replenish fluid reserves
does not exist for {\em G. palpalis} under such circumstances. {\em G.
brevipalpis} is, in turn, considered to be even more hygrophilic than \mbox{{\em
G. palpalis}}. More than half the \mbox{\em G. brevipalpis} pupae still eclode
for pupal environments as dry as 67\% $\mathrm{r.h.}$ (Childs\nocite{Childs1},
2013). Those from the slightly more humid, 71\% $\mathrm{r.h.}$, pupal
environment are resilient enough for general survival through the night until
dawn (Fig. \ref{brevipalpisSurvival}). For those that feed before sunset and
those that eclode into a very humid atmosphere, pupal substrates only minutely
drier might be tolerable and the fact that O.V.I. keep their {\em G.
brevipalpis} colony at 75\% $\mathrm{r.h.}$ (De Beer\nocite{Chantel}, 2013) does
not in any way contradict the aforementioned values. For ideal substrates of
around 85\% $\mathrm{r.h.}$ there is no chance of the newly-eclosed teneral
succumbing to dehydration (Fig. \ref{brevipalpisSurvival}) at room temperature.
The postulated race against time to replenish fluid reserves does not exist for
{\em G. brevipalpis} under those and most other circumstances. Only at the very
driest limit of eclosion does the \mbox{\em G. brevipalpis} teneral succumb to
dehydration, the domain of jeopardy being an extremely narrow band of humidity.
One can therefore state that, for hygrophilic species, the teneral's fate at
room temperature is substantially determined by the conditions which existed in
the pupal environment from which it eclosed. So much so, that it is reasonable
to conclude that, should {\em G. brevipalpis} survive to eclode into atmospheric
conditions which match those of the soil, it will almost certainly survive to
locate a host, without there being any significant prospect of death from
dehydration (Fig. \ref{brevipalpisSurvival}).

Hygrophilic species succumb in response to adverse environmental conditions
earlier, when they are vulnerable, before the teneral stage, and in far greater
numbers. There is no eclosion for a diverse range of pupal conditions, such as
one finds in the case of the mesophilic and xerophilic species. If one considers
this teneral work, in conjunction with the work pertaining to the pupal stage
(Childs\nocite{Childs2}, 2013), a final glance at the data in
Bursell's\nocite{Bursell2} (1959) \mbox{Table II} points to one inevitable
conclusion: The profound revelation that the classification of species as
hygrophilic, mesophilic and xerophilic is largely true only in so much as their
third and fourth instars are and, possibly, the hours shortly before eclosion.
Bursell's\nocite{Bursell2} (1959) \mbox{Table II} reveals that the adult with
the highest, relative transpiration rate is the allegedly-xerophilic \mbox{{\em
G. longipennis}}, whereas the allegedly-hygrophilic, adult {\em G. brevipalpis}
is even more xerophilic than adult \mbox{{\em G. swynnertoni}} and adult {\em G.
pallidipes}! There is, furthermore, very little difference in the relative
losses suffered by the latter three species as adults; or even {\em G.
palpalis}, for that matter. Had Bursell\nocite{Bursell2} (1959), himself, not
remarked on his data, one might find it difficult to believe he hadn't confused
his samples. A simple inspection of Table \ref{modelConversionFactors} reveals
that, while \mbox{{\em G. austeni}} is only a little greater than half the size
of {\em G. morsitans}, it loses the same amount of water. In contrast, {\em G.
brevipalpis} only loses twice as much water as {\em G. morsitans}, this being in
keeping with their relative sizes. An adult {\em G. brevipalpis} might therefore
be considered as xerophilic as \mbox{{\em G. morsitans}} whereas an adult {\em
G. austeni} doesn't stand a chance in dry climates. 

For the first time it becomes clear why {\em G. austeni} is a relatively
sedentary (Childs\nocite{Childs3}, 2010) denizen of low-lying, often coastal,
vleis and estuaries, whereas the highly mobile {\em G. brevipalpis}
(Childs\nocite{Childs3}, 2010) is also associated with drainage lines in the
atmospherically-drier regions of the hinterland. {\em G. austeni} only needs to
larviposit where it feeds while {\em G. brevipalpis} can strike out into drier,
surrounding country to feed, in spite of its pupa requiring the most humid
substrate of all. For the first time it becomes clear why Rogers and
Robinson\nocite{RogersAndRobinson} (2004) found that the predominant variables
in determining the distribution of these two hygrophilic species were NDVI, in
the case of {\em G. brevipalpis}, and elevation in the case of {\em G. austeni}.
As an adult, {\em G. austeni} still requires an humid atmosphere in which to
survive, whereas \mbox{{\em G. brevipalpis}} only requires a humid substrate in
which to larviposit. Hence, the frequent association of \mbox{\em G.
brevipalpis} with drainage lines, remote from their catchment areas and of which
NDVI is the only trace. \mbox{{\em G. austeni}} is always going to be
disadvantaged by its small size, with all the lack of hydrational inertia that a
high surface-area-to-volume ratio implies. {\em G. brevipalpis} is always going
to have an advantage over other species for diametrically opposite arguments. On
the other hand, \mbox{{\em G. austeni}} eclodes much earlier, leaving {\em G.
brevipalpis} to bear the brunt of density-dependent predation and parasitism
(Rogers and Randolph\nocite{RogersAndRandolph1}, 1990). In this way \mbox{{\em
G. austeni}} may compete with \mbox{{\em G. brevipalpis}} in the many
environments which facilitate a sympatric, \mbox{\em G. brevipalpis}-{\em G.
austeni} population (Childs\nocite{Childs3}, 2010).

While it is tempting to assert that {\em G. brevipalpis} and {\em G. austeni}
are the {\em fusca} and {\em morsitans} groups' answer to {\em G. palpalis}-type
environments, they are in many ways more specialized. If {\em G. palpalis} is to
be regarded as a compromise between {\em G. brevipalpis} and {\em G. austeni},
it is unfortunate that there is little, known water-loss data on other members
of the {\em palpalis} group. The steeper terrain on the eastern side of the Rift
Valley generally gives rise to younger geomorphologies characterized by more
clearly differentiated river profiles, consequently to two, more specialized
species. In contrast, much of west Africa is at a relatively low elevation,
giving rise to less energetic rivers, typically associated with a more mature
geomorphology. Some of these rivers are associated with river basins of a low
elevation. Most meander through coastal deltas before terminating in estuaries.
This fact coupled with an higher rainfall, gives rise to soil environments which
generally tend to be more humid than those in the vicinity of the valley
thicket, vleis and estuaries to which {\em G. austeni} is restricted, as an
adult. A drier climate and better drainage on the eastern side of the Rift
Valley ultimately mean that {\em G. austeni} pupae must tolerate drier soils
and, as an adult, never stray too far inland from the coastal plains with their
moist ocean air, in South Africa. A wider distribution of moist soils and
climates in West Africa means the habitats of {\em palpalis}-group flies tend to
be more two-dimensional and less fragmented than those of \mbox{{\em G.
austeni}} and {\em G. brevipalpis}. {\em G. palpalis} can successfully
larviposit closer to where it feeds than {\em G. brevipalpis} sometimes needs
to. {\em G. brevipalpis} must generally larviposit in the immediate vicinity of
water explaining its frequent occurrence in and around riverine forests. The
Rogers and Robinson\nocite{RogersAndRobinson} (2004) study found NDVI to be the
second, most predominant variable in determining the distribution of {\em G.
tachinoides}. This, coupled with a habitat positioned inland from the coast,
might be taken to suggest it is the member of the {\em palpalis} group with the
environment most similar to {\em G. brevipalpis}. {\em G. tachinoides} is,
however, of similar size to {\em G. austeni} and it would appear to flourish in
pupal substrates with an even lower moisture content.

One further, general remark can be made with regard to the limited collection of
results presented. That is the existence of an optimum humidity common to all
species at dawn, as well as a very narrow envelope containing the optima for all
species at the time of eclosion. If pupae eclode in the late afternoon and feed
at sunset, the survival optima for all species lie within an approximately 15\%
$\mathrm{r.h.}$ interval of each other, or less; all other variables being
equal. If pupae eclode in the late afternoon and feed at sunset, the differences
in optimal survival could be consistent with different habitats. If, however,
one compares the lower levels of survival, for example the 50\% level (which
might loosely be interpreted to demarcate survival for a single fly), then the
differences in the ranges of survival conditions for each species are enormous
and there can be no doubt that the species have different habitats. Pupae are,
however, thought to eclode in the evening (Vale et.
al.\nocite{ValeHargroveJordanLangleyAndMews} 1976) and, considering that the
exoskeleton needs a few hours to harden first, one might surmise that the
newly-eclosed fly has to wait through the night, until dawn, to feed. At around
12 hours after eclosion, an optimal, steady humidity (air and soil) for the
survival of all species turns out to exist at around 85\% $\mathrm{r.h.}$ (for a
20\% activity level at 25~$^\circ\mathrm{C}$; Figs.
\ref{morsitansSurvival}--\ref{swynnertoniSurvival}). A slightly lower, 90\%
survival contour also exists for all species, at around the 70\% $\mathrm{r.h.}$
level. A different optimum, common to all species, appears to exist for the case
when the temperature of the pupal substrate and the atmosphere differ. Habitat
and distribution might therefore not be determined by optimal conditions of
survival, instead, by the conditions for which survival is marginal. One problem
with such a conclusion is that it would have been drawn from a limited number of
scenarios in which the humidity of the soil does not differ from that of the
air. 

The model formulated was obviously intended for more ambitious purposes than
mere interpretation and a visualization of data. There are, nonetheless, many
examples which demonstrate that it is an invaluable tool in the interpretation
and visualization of the Bursell\nocite{Bursell1}\nocite{Bursell2} (1958 and
1959) work, to the point of novel biological insight. For example, on assessing
his findings, \mbox{Bursell\nocite{Bursell2} (1959)} concluded that ``for
species like {\em G. palpalis} and {\em G. austeni}, there is reason to suppose
that the soil humidity never drops far below saturation''. By ``never drops
far'' it might be assumed that a drop of somewhere around 10\% $\mathrm{r.h.}$
is indicated. Figures \ref{austeniSurvival} and \ref{palpalisSurvival} suggest
that ``never drops far'' is ideally a drop of around 20\% $\mathrm{r.h.}$ in the
humidity of both the atmosphere and the pupal substrate, however, an average
teneral (with survival loosely defined as being demarcated by the 50\% survival
contour) should still, generally survive at around 65\% $\mathrm{r.h.}$, for
{\em G. palpalis}, and 60\% $\mathrm{r.h.}$, for {\em G. austeni}, not taking
other causes of mortality into account. If the atmosphere is very humid at the
time of eclosion, in the latter case, soils as dry as 55\% $\mathrm{r.h.}$ might
be tolerated (Fig. \ref{bursellsSaturatedSoil}). In the case of {\em G.
austeni}, ``never drops far below saturation'' could be referring to a value as
dry as 55\% $\mathrm{r.h.}$ and the value thus concluded ignores the possibility
that the Bursell\nocite{Bursell1} (1958) pupae may have been of a slightly
inferior quality. Has the model misinterpreted the data? Is there a mistake? No,
the model's prediction is corroborated by, for example, the conditions under
which O.V.I. keep their tsetse colonies. ``Never drops far below saturation''
could, indeed, mean a pupal substrate as dry as 55\% $\mathrm{r.h.}$ in the case
of {\em G. austeni}. Of course, other causes of mortality in the wild would
adjust survival downward and, under these circumstances, the minimum humidity
would need to be revised upward. Of course, if the temperature of the pupal
substrate departs substantially from room temperature then the same would apply.
The point is that {\em G. austeni} can probably survive in soils a lot drier
than are usually claimed. {\em G. austeni} is probably as much limited by
atmospheric humidity as the conditions which prevailed in its pupal environment.
The point is that verbal communication is often too vague. It does not lead to a
precise understanding and can even be misleading. 

One should never lose sight of the fact that the results presented in this work
are projections. They are a substitute which comes a far second to the lab
results which would take a life time of work, at considerable cost. Not everyone
is happy with the predictive potential of such models, phenotypic plasticity
being just one of the reasons. Consider, however, that if a model is able to be
adapted and successfully make certain predictions with respect to other species,
how much more suitable must it be for adaption within a given species. To a
certain extent it can be said that if one is not happy with the model, then one
is really not happy with the data. Another complaint is that there is not enough
data. Of course, there is never enough data and once there is, a model is no
longer necessary. In this case, the acquisition of the outstanding
temperature-dependent data is, however, a priority. There is presently so little
data that the only way the teneral model can be validated is to observe that the
Bursell\nocite{Bursell2} (1959) Figs. 1A, 1C and 1D can be recovered from Fig.
\ref{bursellTeneral}, that the conclusions with regard to the habitats of
\mbox{\em G. austeni} and \mbox{\em G. brevipalpis} appear to be consistent with
the observations of Hendrickx\nocite{Hendrickx} (2007) and that there is no
disparity with the conditions under which O.V.I. keep their tsetse colonies.
There is no way to validate the assumption that the relative transpiration rates
of the different species do not change when no longer at rest and in humid
environments. It is otherwise hoped that the model brings a certain degree of
closure to the question of early mortality due to dehydration in tsetse,
outstanding temperature-dependent data aside. The prognosis for the simplistic
experimental model would seem to be better than expected (given that issues such
as inferior quality pupae, differing puparial durations and the shortage of
statistically significant data can be corrected at some stage). 

This work, in conjunction with \mbox{Childs\nocite{Childs1} (2013)}, claims to
bring about a revision of the conventional wisdom on what truly determines the
classification of species as either hygrophilic, mesophilic or xerophilic. Such
classification has little to do with the tolerance of the eclosed fly to
adversely hot and dry conditions and one possible criticism of the
Bursell\nocite{Bursell2} (1959) conclusion is that it doesn't take the state of
the inherited, pupal reserves into account enough. The plots presented in this
work point to the fact that the sites for larviposition in some species are very
much confined in the dry season, {\em G. brevipalpis} being a case in point. It
must larviposit in close proximity to water. These would be obvious places in
which to concentrate control measures and one immediate application of this
work. Barriers of the type modelled in Childs\nocite{Childs3} (2010) might be
far more efficacious if placed around sites of larviposition than when used for
containment. Unfortunately, in the case of {\em G. brevipalpis}, recent work by
Motloang \nocite{MotloangMasumuVanDenBosscheMajiwaLatif} et. al. (2009) suggests
the species is not a vector of trypanosomiasis and Childs\nocite{Childs3} (2010)
also points to limited evidence that it could, in fact, be in competition with
{\em G. austeni}. {\em G. palpalis} and \mbox{{\em G. austeni}} are most
definitely vectors of trypanosomiasis, although their pupal sites appear not to
be as severely restricted within their adult environments.  

\section{Acknowledgements} 

The author is indebted to Dr. Neil Muller for advising him on statistical
aspects of the probabilistic formulation. The author is indebted to Schalk Schoombie, Johan Meyer and Eleanor \mbox{van der Westhuizen} of the University of the Free State for hosting this work.

\bibliography{teneralH2Oloss}



\end{document}